\documentclass[aps,pre,twocolumn,reprint,amssymb,amsmath,nobibnotes,nofootinbib,superscriptaddress,floatfix]{revtex4}
\usepackage[dvips]{graphicx}
\usepackage[hidelinks]{hyperref}
\usepackage{epsfig}
\usepackage{framed}
\usepackage{color}
\usepackage{xcolor}
\usepackage{textcomp}
\usepackage{setspace}
\usepackage{morefloats}
\usepackage[T1]{fontenc}
\usepackage[utf8]{inputenc}
\usepackage{cleveref}
\usepackage{ulem}
\usepackage{float}
\usepackage{bm}
\usepackage{comment}
\usepackage{titlesec}
\usepackage{titletoc}

\usepackage[scr=boondox]{mathalfa}
\allowdisplaybreaks[2]

\usepackage{array}
\newcommand{\PreserveBackslash}[1]{\let\temp=\\#1\let\\=\temp}
\newcolumntype{C}[1]{>{\PreserveBackslash\centering}p{#1}}

\titleformat{\section}[block]{\scshape\centering\bfseries}{\thesection}{1pt}{.\enspace}

\begin{document}

\title{Emergent complex phases in a discrete flocking model with reciprocal and non-reciprocal interactions}

\author{Matthieu Mangeat}
\email{mangeat@lusi.uni-sb.de}
\affiliation{Center for Biophysics \& Department for Theoretical Physics, Saarland University, 66123 Saarbr{\"u}cken, Germany.}

\author{Swarnajit Chatterjee}
\email{swarnajit.chatterjee@uni-saarland.de}
\affiliation{Center for Biophysics \& Department for Theoretical Physics, Saarland University, 66123 Saarbr{\"u}cken, Germany.}

\author{Jae Dong Noh}
\email{jdnoh@uos.ac.kr}
\affiliation{Department of Physics, University of Seoul, Seoul 02504, Korea.}

\author{Heiko Rieger}
\email{heiko.rieger@uni-saarland.de}
\affiliation{Center for Biophysics \& Department for Theoretical Physics, Saarland University, 66123 Saarbr{\"u}cken, Germany.}

\begin{abstract} 
There is growing interest in multi-species active matter systems with reciprocal and non-reciprocal interactions. While such interactions have been explored in continuous symmetry models, less is known about multi-species discrete-symmetry systems. To address this, we study the two-species active Ising model (TSAIM), a discrete counterpart of the two-species Vicsek model. Our investigation explores both inter-species reciprocal and non-reciprocal interactions, along with the possibility of species interconversion. In the reciprocal TSAIM, we observe the emergence of a high-density parallel flocking state, a feature not seen in previous flocking models. With species interconversion, the TSAIM corresponds to an active extension of the Ashkin-Teller model and exhibits rich state diagrams. In the non-reciprocal TSAIM, a run-and-chase dynamics emerge. We also find that the system is metastable due to droplet excitation and exhibits spontaneous motility-induced interface pinning. A hydrodynamic theory validates our numerical simulations and confirms the phase diagrams.
\end{abstract}
%\keywords{Flocking; Active Matter; Metastability}

\maketitle

%%%%%%%%%%%%%%%%%%%%
%%% INTRODUCTION %%%
%%%%%%%%%%%%%%%%%%%%

\section{Introduction}
\label{sec1}
Active matter is a class of natural or synthetic non-equilibrium systems composed of many agents that consume energy to move or exert mechanical forces. Over the past two decades, intensive research has established active matter as a significant field of study~\cite{marchetti2013,ramaswamy,magistris2015,volpe2016,shaebani2020,marchetti2022}, with assemblies of active particles exhibiting complex behaviors and collective effects, such as the emergence of large, ordered clusters known as flocks. Flocking plays a significant role in a wide range of systems across disciplines including physics, biology, ecology, social sciences, and neurosciences~\cite{strogatz} and is an out-of-equilibrium phenomenon abundantly observed in nature~\cite{marchetti2013,volpe2016,shaebani2020,marchetti2022}.

A widely studied computational model for flocking is the celebrated Vicsek model (VM)~\cite{Viscek1995,ginelli2016}. In this model, point particles with rotational symmetry tend to align with their neighbors while moving at a fixed speed. This alignment is not perfect as particles make error which are modeled as a stochastic noise. Although the VM shows a transition from a disordered (low-density, high-noise) to an ordered (high-density, low-noise) phase, Solon et al.~\cite{solon2015a,solon2015b} demonstrated that this is better understood as a liquid-gas transition with a phase-separated coexistence region for intermediate noise and density. Remarkably, despite its continuous symmetry, the VM exhibits true long-range order in two dimensions, seemingly violating the Mermin-Wagner theorem, which prohibits the spontaneous breaking of continuous symmetry in two-dimensional systems in thermal equilibrium. However, as Toner and Tu~\cite{TT1995,TT1998,TT2012} showed through their hydrodynamic theory, this apparent violation arises because the moving flock operates far from equilibrium. The non-equilibrium nature of the system, driven by the motion of particles, allows for the emergence of macroscopic order and the breaking of rotational invariance.

Subsequently, the active Ising model (AIM)~\cite{AIMprl,AIMpre} was introduced which replaces the continuous rotational symmetry of the VM with a discrete symmetry, while preserving the essential physics of the VM. The AIM exhibits three steady-states similar to the VM: disordered gas at high noise and low densities, polar liquid at low noise and high densities, and a phase-separated liquid-gas coexistence region at intermediate densities and temperatures. The key distinction between the VM and the AIM is that the former exhibits giant density fluctuations leading to microphase separation of the coexistence region, while the latter shows normal density fluctuations resulting in bulk phase separation. Recent studies on the $q$-state active Potts model (APM)~\cite{APMpre,APMepl,APMjamming} and the $q$-state active clock model (ACM)~\cite{ACMepl,ACMprl} have provided further insights into flocking transitions. These models have emerged as more generalized frameworks for flocking bridging the VM and the AIM.

Due to their inherent out-of-equilibrium nature, active systems exhibit long-range order (LRO)~\cite{Viscek1995} and survive spin wave fluctuations~\cite{TT1995,TT1998,TT2012}. Consequently, it was thought that polar-ordered phases in the active matter are generally robust to fluctuations. In addition, studies of these flocking models are typically conducted in idealized settings, assuming perfectly identical particles in an infinite and homogeneous environment. However, real flocks are more complex, and recent studies have revealed that even a single small obstacle or artificially excited droplet within the ordered phase can destabilize it, as observed in the VM~\cite{metaVM,DVM}. This contrasts sharply with passive systems, where such a small perturbation typically only induces a local effect. Furthermore, these perturbations can emerge spontaneously and this spontaneous fluctuation could similarly destabilize the ordered phase, as recently observed in the constant-density Toner-Tu flocks~\cite{metaTT} and the AIM~\cite{metaAIM,mip}, where ordered flocks become metastable over large time scales, eventually transitioning to a disordered state.

Additionally, the models of self-propelled particles mostly focus on homogeneous systems where every
agent has exactly the same dynamical properties and follows the same ``rules of engagement''. However, heterogeneity is ubiquitous in nature and complex systems are typically heterogeneous as
individuals have disparate characteristics and vary in their properties~\cite{hetero1,hetero2,hetero3}. In particular, many biological systems that show flocking involve self-propelled particles with heterogeneous interactions which can significantly impact system dynamics, as seen in studies of mixed bacterial populations~\cite{mixed-species-Ariel1,population-segregation,mixed-species-Ariel2}. Theoretically, various aspects of heterogeneous systems of self-propelled agents have been investigated. For example, alignment interactions in a binary mixture of self-propelled particles have shown that different interaction potentials can lead to parallel, antiparallel, or perpendicular alignment, resulting in diverse collective motion patterns~\cite{menzel}. Other investigations have explored particles with varying velocities~\cite{mixed-velocities1,mixed-velocities2}, noise sensitivity~\cite{hetero3,mixed-noise}, sensitivity to external cues~\cite{mixed-external}, and particle-to-particle interactions~\cite{mixed-interactions1,mixed-interactions2,mixed-interactions3}. Different self-propelled particle species were also analyzed in predator-prey scenarios~\cite{pp1,pp2} and in the context of reciprocal, e.g. the two-species Vicsek model (TSVM)~\cite{TSVM}, as well as non-reciprocal interactions~\cite{fruchart,vitteliNRAIM}.

The TSVM~\cite{TSVM}, which is a two-species extension of the VM with reciprocal antiferromagnetic interspecies interactions, exhibits two primary steady states describing the collective motion: the anti-parallel flocking (APF) state, where the two species form bands moving in opposite directions, and the parallel flocking (PF) state, where the bands travel in the same direction. In the low-density and high-noise part of the coexistence region, PF and APF states perform fluctuation-induced stochastic transitions from one to the other where the transition frequency decreases with increasing system size. Furthermore, the PF state vanishes at high densities and low noises, leaving the APF state as the only ordered {\it liquid} phase.

Finally, an increasing number of recent studies involving non-equilibrium systems have focused on how non-reciprocal interactions in active matter affect non-equilibrium phase transitions and drive the emergence of states or patterns in active matter~\cite{marchetti2022, lowen2015}. A non-reciprocal interaction violates Newton's third law ``actio=reactio'' and leads to frustration between two elements due to their opposing objectives. In soft and active matter systems, non-reciprocity arises when interparticle forces are mediated by a non-equilibrium environment, leading to the emergence of novel self-organized states dependent on time~\cite{fruchart,Marchetti2020,golas2020,vitteliNRAIM,NRrobot2024,NRklappPRE,NRklappPRL}. More prominent, if not ubiquitous, are non-reciprocal interactions in active and living systems that break detailed balance at the microscale, from social forces~\cite{strogatz2011} and neural networks~\cite{hansel2000,kopell2003} to antagonistic interspecies interactions in bacteria~\cite{xiong2020}, cells~\cite{mayor2013} and predator-prey systems~\cite{meredith2020}. In contrast to equilibrium systems governed by Newton’s third law, non-reciprocal systems are generally considered to be out of equilibrium~\cite{klapp2020} and therefore non-reciprocal interactions are associated with a gain or a loss of energy. Although the incorporation of non-reciprocal interactions is not required to capture flocking behavior, there is a recent push toward understanding the effects of non-reciprocal interaction on the phase behavior of flocking objects~\cite{Durve2016,Durve2018,fruchart,vitteliNRAIM}.

In this paper, we investigate a two-species variant of the active Ising model (AIM)~\cite{AIMprl,AIMpre}, namely the two-species active Ising model (TSAIM), which serves as a discrete-symmetry counterpart to the continuous-symmetry TSVM~\cite{TSVM}. In the TSAIM, self-propelled particles from two distinct {\it species} A and B undergo biased diffusion in two dimensions along with local alignment. In this context, we have mainly considered three different scenarios:\\
(a) reciprocal TSAIM with conserved species where a particle aligns with particles of the same species and anti-aligns with particles of the other species but the population of each species remains conserved, corresponding to the discrete counterpart of the TSVM~\cite{TSVM};\\
(b) reciprocal TSAIM with non-conserved species where the alignment protocol follows (a) but a particle of one species can convert to the other species, representing an active extension of the Ashkin-Teller model, equivalent to two coupled Ising-like subsystems~\cite{AT1,AT2}, and could be interpreted as a binary voter model~\cite{voter1,voter2} for self-propelled agents;\\
(c) non-reciprocal TSAIM (NRTSAIM) with conserved species where particles of different species interact in a non-reciprocal manner: A particles tend to align with B, while the B particles tend to anti-align with A, acting as a natural extension of the AIM mimicking a predator-prey model~\cite{pp2}, where species A plays the role of the predator and species B plays the role of the prey.

\section{Results}

%%%%%%%%%%%%%%%%%%%%%%%%%%%%%%%%%
%%% MODEL: MICRO UPDATE RULES %%%
%%%%%%%%%%%%%%%%%%%%%%%%%%%%%%%%%

\subsection{Microscopic model}

We consider an ensemble of $N$ particles in a periodic two-dimensional square lattice of size $L_x \times L_y$. The average particle density is $\rho_0 = N / L_x L_y$. The $j^{\rm th}$ particle on site $i$ is equipped with a spin-orientation $\sigma_i^j = \pm 1$ which determines the biased hopping with self-propulsion $\varepsilon$ via the rate:
\begin{equation}
W_{\rm hop}(\sigma,{\bf p}) = D\left(1+\theta \varepsilon |{\bf p} \cdot {\bf e_x}| +\sigma \varepsilon {\bf p} \cdot {\bf e_x} \right),\label{eqhop}
\end{equation}
with $\theta \in [0,1]$, i.e. a rate $W_+=D[1+(\theta+1)\varepsilon]$ in the favored direction (${\bf p}=\sigma {\bf e_x}$), a rate $W_-=D[1+(\theta-1)\varepsilon]$ in the unfavored direction (${\bf p} = -\sigma {\bf e_x}$), and a constant rate $D$ in the upward and downward directions (${\bf p} = \pm {\bf e_y}$). The parameter $0 \le \varepsilon \le 1/(1-\theta)$ controls the asymmetry between the non-motile limit $\varepsilon = 0$ ($W_+=W_-=D$) and the limit where the particle never jumps to the unfavored direction $\varepsilon = 1/(1-\theta)$ ($W_-=0$). On average, a particle drifts with speed $v=2D\varepsilon$ in the direction set by the sign of its spin orientation, and diffuses with diffusion constant $D_{xx} = D+\theta v/2$ and $D_{yy}=D$ along $x$ and $y$ directions, respectively. The total hopping rate thus becomes $4D + \theta v$, and the limiting values in $\varepsilon$ give the inequality $v \le 2D/(1-\theta)$. The case $\theta=0$, i.e. $W_\pm=D(1\pm \varepsilon)$, will be used to derive the main results, as in Ref.~\cite{AIMprl,AIMpre}, while the case $\theta=1$, i.e. $W_-=D$ and $W_+=D+v$, will be used to study the stability of the liquid state, as in Refs.~\cite{mip,metaAIM}, since the diffusion $D$ can be made small in comparison to the velocity $v$.

The particle is further equipped with a species degree of freedom $s_i^j = \pm 1$ which defines the particle species ($s_A=1$ and $s_B=-1$). The number of particles on site $i$, with spin orientation $\sigma$ and species spin $s$ is denoted by $n_{s,i}^\sigma$. No restriction is applied to the number of particles of species $s$ on site $i$: $\rho_{s,i} = \sum_{\sigma} n_{s,i}^\sigma$ and to the total number of particles on site $i$: $\rho_i = \rho_{A,i} + \rho_{B,i}$. The local magnetization of species $s$ on site $i$ is defined by: $m_{s,i} = \sum_{j\in s} \sigma_i^j = n_{s,i}^+ - n_{s,i}^-$.

Analogous to the one-species AIM~\cite{AIMprl,AIMpre}, we consider the following flipping rate for a particle with spin $\sigma$ and species $s$ on site $i$, for an inverse temperature $\beta$:
\begin{equation}
W_{\rm flip}^{(1)}(\sigma \to -\sigma) = \gamma_1 \exp\left(- \frac{2 \beta}{\rho_i} \sigma J_{ss'} m_{s',i} \right),\label{wflip1}
\end{equation}
where the Einstein notation is used: $J_{ss'} m_{s',i} =  J_{sA} m_{A,i} +  J_{sB} m_{B,i}$. Here, $J_{ss'}$ represents the coupling constant of species $s'$ acting on species $s$.

First, we consider reciprocal interactions between the two species. Similar to the TSVM~\cite{TSVM}, the coupling constant is given by $J_{ss'} = J_1 s s'$ (with $J_1>0$), which is positive when the two species are the same (ferromagnetic interaction) and negative when the species are different (anti-ferromagnetic interaction). We can now define the following two key order parameters~\cite{TSVM}: the total local magnetization
\begin{equation}
v_{s,i} = \sum_j \sigma_i^j = m_{A,i} + m_{B,i},
\end{equation}
defining the average propulsion direction at site $i$, and the local magnetization difference
\begin{equation}
v_{a,i} = \sum_j s_i^j \sigma_i^j = m_{A,i} - m_{B,i},
\end{equation}
playing the role of the order parameter of the ferromagnetic interaction. From Eq.~\eqref{wflip1}, the flipping rate for the spin-orientation in the reciprocal TSAIM then becomes:
\begin{equation}
W_{\rm flip}^{(1)}(\sigma \to -\sigma) = \gamma_1 \exp\left(- \frac{2 \beta J_1}{\rho_i} s \sigma v_{a,i}  \right).\label{wflip1bis}
\end{equation}
Note that, this expression is similar to the one-species flipping rate~\cite{AIMprl,AIMpre}, with $v_{a,i}$ replacing the one-species magnetization.

In the case of non-reciprocal interactions between species, we primarily focus on the scenario where $J_{AA} = J_{BB} = J_1$ and $J_{AB} = - J_{BA} = J_{\rm NR} \le J_1$. From Eq.~\eqref{wflip1}, the flipping rate for the spin-orientation in the non-reciprocal TSAIM is defined as the following:
\begin{equation}
W_{\rm flip}^{(1)}(\sigma \to -\sigma) = \gamma_1 \exp\left(- \frac{2 \beta J_1}{\rho_i} \sigma \mu_{s,i} \right),\label{NRflip}
\end{equation}
where $\mu_{s,i} = m_{s,i} + s (J_{\rm NR}/ J_1) m_{-s,i} $ plays the role of order parameter for species $s$. We also investigate a more general coupling constant such that $- J_1 \le J_{BA} \le 0 \le J_{AB} \le J_1$, where species A interacts ferromagnetically with species B, while species B has an anti-ferromagnetic interaction with species A.

Additionally, the spins $s_i^j$ defining the particle species may interact locally on site $i$, via an Ising interaction independent of the spin-orientation of the particles. As for the one-species AIM, we consider the following flipping rate for the species-spin in the TSAIM:
\begin{equation}
W_{\rm flip}^{(2)}(s \to -s) = \gamma_2 \exp\left(- \frac{2 \beta J_2}{\rho_i} s m_i \right), \label{wflip2}
\end{equation}
where the coupling constant $J_2$ of species-species interactions is positive, and $m_i = \sum_{j} s_i^j = \rho_{A,i} - \rho_{B,i}$ is the species magnetization on site $i$. We define $m_0=(N_A-N_B)/L_xL_y$ as the average species magnetization at the initial time.

The reciprocal TSAIM with species flip ($\gamma_2 \ne 0$) is a natural {\it active} extension of the well-known Ashkin-Teller (AT) model~\cite{AT1}, in which four components interact, equivalent to two coupled Ising-like subsystems (here coupling spins $\sigma$ and species $s$)~\cite{AT2}. The order parameters $\langle v_s \rangle \sim \langle \sigma \rangle$, $\langle m \rangle \sim \langle s \rangle$, and $\langle v_a \rangle \sim \langle \sigma s \rangle$ measure the relative ordering between the spins and species~\cite{baxter1,baxter2}.

It is crucial to emphasize that three separate magnetizations are explicitly defined at each site $i$ in this model: (i) the magnetization of species A, denoted $m_{A,i} = \sum_{j\in A} \sigma_i^j$, (ii) the magnetization of species B, denoted $m_{B,i} = \sum_{j\in B} \sigma_i^j$, and (iii) the species magnetization, denoted $m_i = \sum_{j} s_i^j$. The state of the site $i$ is completely defined by these three magnetizations and the total density $\rho_i$. In the following, we will consider $\rho_i \equiv \rho(x_i,y_i)$, $m_i \equiv m(x_i,y_i)$, $\rho_{s,i} \equiv \rho_s(x_i,y_i)$, and $m_{s,i} \equiv m_s(x_i,y_i)$, as equivalent expressions where $(x_i,y_i)$ are the coordinates of the site $i$.

Without any loss of generality, we take $\gamma_1 = 1$, and we denote $\beta_1 \equiv \beta J_1 = T_1^{-1}$ and $\beta_2 \equiv \beta J_2 = T_2^{-1}$ playing the role of two effective inverse temperatures for the spin-spin and species-species interactions, respectively. Simulations are performed for several control parameters: the average density $\rho_0$, the self-propulsion parameter $\varepsilon$, and the different effective temperatures or coupling constants ($T_1$ and $T_2$ in the case of reciprocal interactions, $J_{AB}$, $J_{BA}$ and $T_1 \equiv \beta_1^{-1}$ in case of non-reciprocal interactions).

%%%%%%%%%%%%%%%%%%%%%%%%%%%%%%
%%% HYDRODYNAMIC EQUATIONS %%%
%%%%%%%%%%%%%%%%%%%%%%%%%%%%%%

\subsection{Hydrodynamic equations}

In this subsection, we will derive the hydrodynamic equations from the microscopic update rules. From the average particle density $\rho_s^{\sigma}({\bf x};t) \equiv \langle n_s^\sigma({\bf x};t) \rangle$ in state $\sigma$ and species $s$, at position ${\bf x}$ and time $t$, we define the particle density $\rho_s({\bf x};t) = \sum_{\sigma} \rho_s^{\sigma}({\bf x};t)$ and the magnetization $m_s({\bf x};t) = \sum_{\sigma} \sigma \rho_s^{\sigma}({\bf x};t)$ of species $s$. These four functions ($\rho_A$, $\rho_B$, $m_A$, and $m_B$) determine completely the spin and species states at the position ${\bf x}$ and the time $t$.

In Supplementary Note~1, we derive the hydrodynamic equations for the reciprocal TSAIM. From the symmetries of the problem, we define the density $\rho = \sum_{s,\sigma} \rho_s^{\sigma} = \rho_A+\rho_B$, the species magnetization $m = \sum_{s,\sigma} s \rho_s^{\sigma} = \rho_A-\rho_B$, and the order parameters $v_s = \sum_{s,\sigma} \sigma \rho_s^{\sigma} = m_A+m_B$, and $v_a = \sum_{s,\sigma} s\sigma \rho_s^{\sigma} = m_A-m_B$. Since the classical mean-field approximation fails to exhibit phase-separated profiles, we follow the refined mean-field approximation used for the one-species AIM~\cite{AIMprl,AIMpre}, which implies here to take $m$ and $v_a$ as independent Gaussian variables with a variance linear in the density $\rho$: $\sigma_m^2 = \alpha_m \rho$ and $\sigma_a^2 = \alpha_a \rho$, respectively. The hydrodynamic equations, averaged over the Gaussian variables, for the reciprocal TSAIM read
\begin{align}
\partial_t \rho &=  D_{xx} \partial_{x}^2 \rho + D_{yy} \partial_{y}^2 \rho - v \partial_{x} v_s, \label{recEq1} \\
\partial_t m &=   D_{xx} \partial_{x}^2 m + D_{yy} \partial_{y}^2 m - v \partial_{x} v_a\nonumber\\
+& 2\overline{\gamma_2} \left[ \left(\rho- \frac{r_2}{2\beta_2} \right) \sinh \frac{2\beta_2 m}{\rho} - m \cosh \frac{2\beta_2 m}{\rho}  \right], \label{recEq2} \\
\partial_t v_s &=   D_{xx} \partial_{x}^2 v_s + D_{yy} \partial_{y}^2 v_s - v \partial_{x} \rho \nonumber\\
+& 2 \overline{\gamma_1} \left[ m \sinh \frac{2\beta_1 v_a}{\rho} - v_s \cosh \frac{2\beta_1 v_a}{\rho} \right], \label{recEq3}\\
\partial_t v_a &=   D_{xx} \partial_{x}^2 v_a + D_{yy} \partial_{y}^2 v_a - v \partial_{x} m \nonumber\\
+& 2 \overline{\gamma_1} \left[ \left(\rho- \frac{r_1}{2\beta_1} \right) \sinh \frac{2\beta_1 v_a}{\rho} - v_a \cosh \frac{2\beta_1 v_a}{\rho} \right] \nonumber \\
+& 2 \overline{\gamma_2} \left[ v_s \sinh \frac{2\beta_2 m}{\rho}  - v_a \cosh \frac{2\beta_2 m}{\rho} \right], \label{recEq4}
\end{align}
with diffusion constants $D_{xx}=D+\theta v/2$, and $D_{yy}=D$, the self-propulsion velocity $v=2D\varepsilon$, $\overline{\gamma_i}=\gamma_i \exp(r_i/2\rho)$, $r_1=(2\beta_1)^2 \alpha_a$, and $r_2=(2\beta_2)^2 \alpha_m$. Note that $r_1$ and $r_2$ are considered as two new parameters for the hydrodynamic theory, and $r_1=r_2=0$ corresponds to the classical mean-field approximation. We will consider in this paper only the case $D=1$ and $\theta=0$ (i.e. $D_{xx}=D_{yy}=1$ and $v=2\varepsilon$).

In Supplementary Note~2, we derive the hydrodynamic equations for the non-reciprocal TSAIM. As for reciprocal interactions, we follow the refined mean-field approximation consisting here to take the magnetizations $m_s$ as independent Gaussian variables with a variance linear in the density $\rho$: $\sigma_s^2 = \alpha_s \rho$. The hydrodynamic equations, averaged for the Gaussian variables, for the non-reciprocal TSAIM read:
\begin{align}
\partial_t \rho_s &=   D_{xx} \partial_{x}^2 \rho_s + D_{yy} \partial_{y}^2 \rho_s - v \partial_{x} m_s, \label{recEq5} \\
\partial_t m_s &=   D_{xx} \partial_{x}^2 m_s + D_{yy} \partial_{y}^2 m_s - v \partial_{x} \rho_s \nonumber \\
+& 2 \overline{\gamma_s} \left[ \left( \rho_s - \frac{r_1}{2 \beta_1} \right)\sinh  \frac{2\beta_1 \mu_s}{\rho}  - m_s \cosh \frac{2\beta_1 \mu_s}{\rho} \right], \label{recEq6}
\end{align}
with diffusion constants $D_{xx}=D+\theta v/2$, and $D_{yy}=D$, the self-propulsion velocity $v=2D\varepsilon$, $\mu_s = m_s + (J_{s,-s}/J_1) m_{-s}$, $\overline{\gamma_s} = \gamma_1 \exp[(r_1+r_{s,-s})/2\rho]$, and $r_{ss'} = (2\beta J_{ss'})^2 \alpha_{s'} = (J_{ss'}/J_1)^2 r_1$, where $r_1$ can be considered as a new parameter in the context of the hydrodynamic theory. We will consider in this paper only the case $D=1$ and $\theta=0$ (i.e. $D_{xx}=D_{yy}=1$ and $v=2\varepsilon$).

%%%%%%%%%%%%%%%%%%%%%%%%%%%%%%%%%%%%%
%%% RESULTS: WITHOUT SPECIES FLIP %%%
%%%%%%%%%%%%%%%%%%%%%%%%%%%%%%%%%%%%%

\subsection{Reciprocal interactions without species flip ($\gamma_2=0$)}

In this subsection, we present the results of the reciprocal TSAIM without species flip ($\gamma_2=0$) for an equal population of the two species: $N_A=N_B=N/2$, i.e. $m_0=0$, which will stay constant over time. We aim to compare the results of this model, a model with discrete symmetry, with those obtained for the TSVM~\cite{TSVM}, a model characterized by continuous symmetry.

\begin{figure*}[t]
\begin{center}
\includegraphics[width=\textwidth]{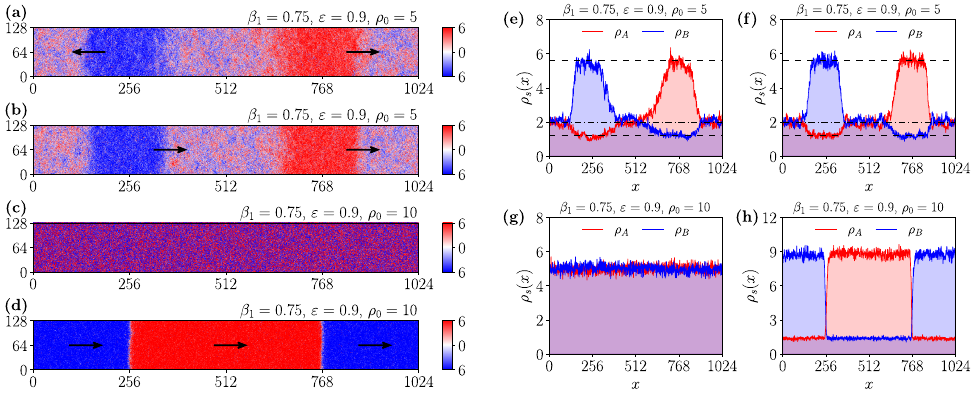}
\caption{{\it Steady states for reciprocal interactions without species flip.} (a--d)~Density snapshots for $\beta_1=0.75$, $\varepsilon=0.9$ and increasing density, in a $1024\times 128$ domain, showing different states: coexistence (a)~APF and (b)~PF states for $\rho_0=5$, (c)~liquid APF state, and (d)~HDPF state for $\rho_0=10$. The densities of A and B species are represented with red and blue colors, respectively. At each site, the color corresponding to the species with the higher density is displayed. High-density bands move in the direction indicated by the arrows. A movie (\texttt{movie 1}) of the same can be found at Ref.~\cite{movies}. (e--h) Density profiles (red/blue represent species A/B, respectively) obtained from the density snapshots shown in (a--d) by integrating along the $y$-axis. The dash-dotted line represents the density in the gas region, while the densities of the two species in the liquid phase are represented by dashed lines, as a guide to the eyes.\label{fig1}}
\end{center}
\end{figure*}

{\bf Steady-state profiles and state diagrams of the TSAIM without species flip.} In this paragraph, we present the results for a hopping rate, given by Eq.~\eqref{eqhop}, with $\theta=0$ and $D=1$. Figs.~\ref{fig1}(a--d) show the steady-state density snapshots obtained from numerical simulations for increasing density starting from a semi-ordered PF configuration. At low densities ($\rho_0<3.98$), the particles of both species are disordered, forming a {\it gas} phase. At higher densities ($3.98 < \rho_0 < 6.82$), particles of both species exhibit flocking behavior in a moving transverse band, forming phase-separated density profiles, characterized by a band of each species moving together on a gaseous background composed of both A and B particles. These two bands are macrophase separated (i.e. only one band is observed in the steady state for each species, contrary to the TSVM), and define the {\it coexistence} region. The A- and B-band can either move in opposite directions, denoted as anti-parallel flocking (APF) [Fig.~\ref{fig1}(a)], or in the same direction, denoted as parallel flocking (PF) [Fig.~\ref{fig1}(b)]. In both the bands, A-particles are anti-aligned with B-particles due to the antiferromagnetic interaction, yielding an opposite motion of B-particles inside the A-band, and vice versa. At even higher densities ($\rho_0>6.82$), the TSAIM exhibits a {\it liquid} state, though the morphology of this state is not unique. The homogeneous liquid phase can emerge where the densities of species A and species B are spatially uniformly distributed in an APF state, as shown in Fig.~\ref{fig1}(c), similar to the liquid phase observed in the TSVM. Additionally, the system can also display a stable high-density PF (HDPF) state, characterized by two parallel flocking bands occupying half of the domain ($L_x/2 \times L_y$) [Fig.~\ref{fig1}(d)]. Unlike in the TSAIM, this HDPF state is unstable in the TSVM~\cite{TSVM}, since the continuous symmetry implies giant number fluctuations~\cite{solon2015a}, and fluctuations in the band orientation are also in favor of the APF state.

\begin{figure}[t]
\begin{center}
\includegraphics[width=\columnwidth]{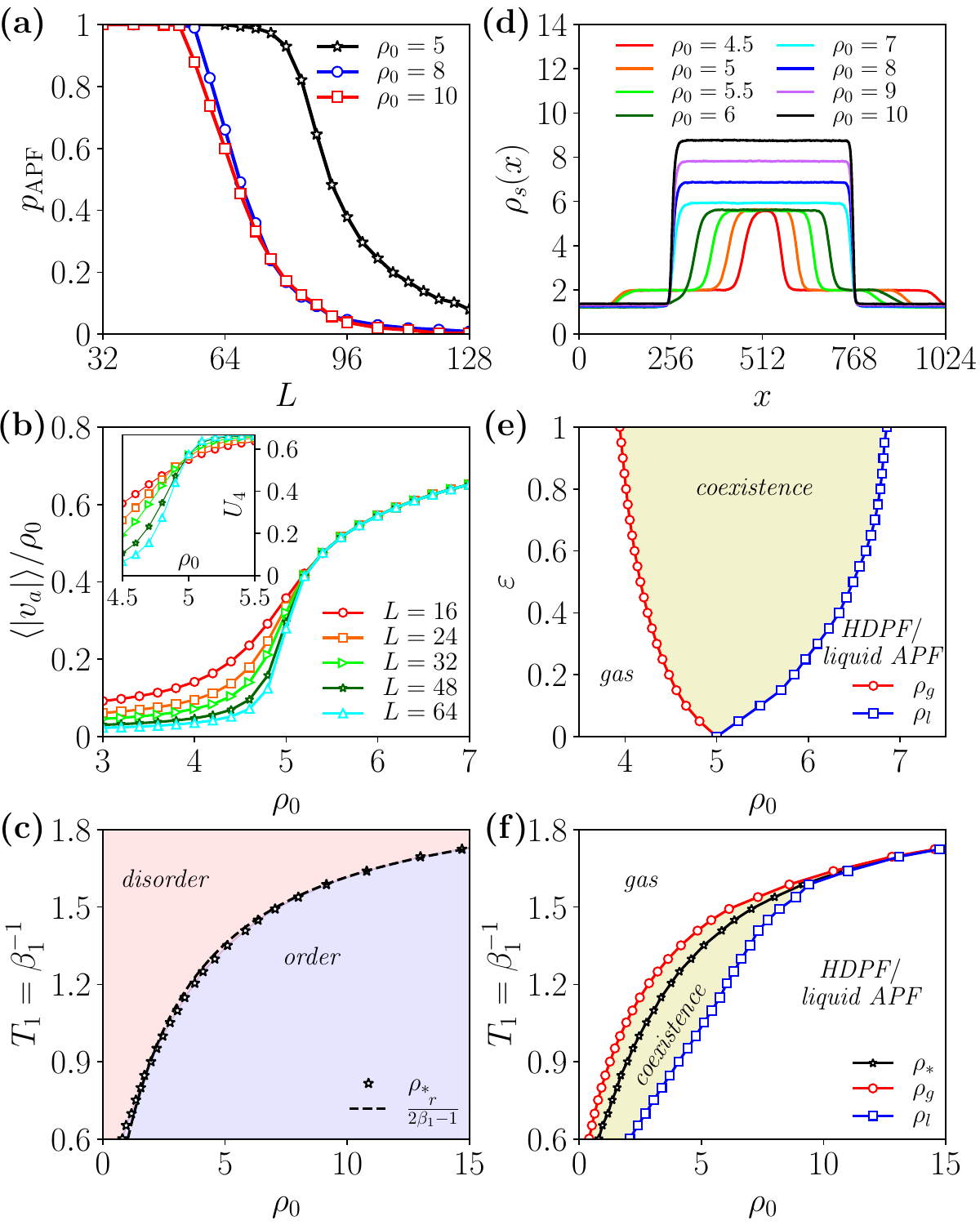}
\caption{{\it Profiles and state diagrams for reciprocal interactions without species flip.} (a)~Probability for a steady state in the APF state as a function of the system size, starting from disordered initial configurations, calculated over $5000$ ensembles for $\beta_1=0.75$, $\varepsilon=0.9$ and several densities $\rho_0$. (b)~Mean value and Binder cumulant (inset) of the order parameter $|v_a|$ vs $\rho_0$ without activity ($\varepsilon=0$) for $\beta_1=0.75$ and several system sizes. The transition occurs at density $\rho_* = 4.95$. (c)~State diagram for $\varepsilon=0$. $\rho_0 < \rho_*(T_1)$: disordered state (red), $\rho_0 > \rho_*(T_1)$: ordered state (blue). (d)~Steady-state density profiles (PF state) for $\beta_1=0.75$ and $\varepsilon=0.9$ in a $1024\times 128$ domain, for various densities. The data are averaged over $1000$ times. (e)~Velocity-density state diagram for $\beta_1=0.75$. (f)~Temperature-density state diagram for $\varepsilon=0.9$. \label{fig2}}
\end{center}
\end{figure}

Figs.~\ref{fig1}(e--h) show instantaneous steady-state density profiles corresponding to Figs.~\ref{fig1}(a--d). The phase-separated density profiles, shown in Figs.~\ref{fig1}(e--f) and characterizing an APF and PF state, respectively, exhibit three distinct regions for each species: a maximum density in its own liquid band, an intermediate density in the gas phase outside both bands, and a minimum density in the liquid band of the other species. Due to the antiferromagnetic interaction between the two species, when a liquid band of species $s$ moves to the right (left), with a positive (negative) magnetization, the particles of species $-s$ inside that band go to the left (right) with a negative (positive) magnetization. As the density increases, with an APF initial condition, both species exhibit homogeneous profiles around their average density, indicative of an APF state [Fig.~\ref{fig1}(g)]. Starting from a PF initial condition, however, the system forms individual high-density bands for each species in the HDPF state, together spanning the $x$-axis [Fig.~\ref{fig2}(h)]. In the HDPF state, no gas phase is present, leading to two distinct regions in the density profiles: a maximum density within the own band of each species and a minimum density within the band of the other species.

%\begin{figure}[t]
%\begin{center}
%\includegraphics[width=\columnwidth]{fig3.pdf}
%\caption{(color online) Time evolution of the reciprocal TSAIM without species flip ($\gamma_2=0$) starting from a disordered initial configuration to an HDPF state for $\beta_1=0.75$, $\varepsilon=0.9$, $\rho_0=8$ in a $512\times 512$ domain (colormap as in Fig.~\ref{fig1}). \label{fig3}}
%\end{center}
%\end{figure}

%To understand the existence of this HDPF state, Fig.~\ref{fig3} shows the time evolution of the TSAIM starting from a disordered configuration, for which the steady state is in the HDPF state. We initially observe the nucleation of several flocks of species A and B moving either to the left or right [Figs.~\ref{fig3}(a--c)]. As time progresses, these flocks grow, with pairs of parallel flocks of species A and B being favored [Fig.~\ref{fig3}(d)]. This results in the formation of multiple PF bands arranged in layers along the $y$-axis [Fig.~\ref{fig3}(e)], with alternating band motions to the left and right. At longer times, one of the layers becomes dominant, in this case moving to the right [Fig.~\ref{fig3}(f)]. Conversely, in the TSVM~\cite{TSVM}, nucleating flocks are randomly oriented, not limited to anti-parallel or parallel flocking because of the continuous symmetry of spin orientation, and the growth of these nuclei does not favor the PF state over the APF state at large times.

The emergence of this HDPF state involves the nucleation and growth of several flocks of both species, eventually leading to layered polar bands with a dominant direction of motion. See Supplementary Note~3 for a detailed description of this process through the time evolution of the TSAIM starting from a disordered configuration.

Fig.~\ref{fig2}(a) shows the probability $p_{\rm APF}$ of having an APF steady state, rather than a PF steady state, starting from a disordered initial configuration, as a function of the domain size $L$, calculated over $5000$ different initial configurations and for several densities $\rho_0$. A transition occurs between small system sizes, where the system exhibits an APF steady state for all realizations, and larger system sizes, where a PF steady state dominates. The probability $p_{\rm APF}$ becomes smaller than 1 (i.e., at least one configuration goes to a PF state) for $L > L_c$, with $L_c \sim 55$ for $\rho_0\in \{8,10\}$ and $L_c \sim 80$ for $\rho_0=5$, leading to HDPF and phase-separated PF steady states, respectively. This suggests that PF states are more {\it stable} than APF states in the hydrodynamic limit ($L \to \infty$) when starting from a disordered initial condition. It is important to note that, for semi-ordered initial conditions, the final TSAIM steady state is determined by the initial configuration, a PF (APF) initial condition always leads to a PF (APF) state. Indeed, we do not observe any fluctuation-induced stochastic switching between the APF and PF states in the TSAIM coexistence region, as seen in the TSVM~\cite{TSVM}, since the density fluctuations in the TSAIM are normal, similar to the one-species AIM~\cite{AIMprl,AIMpre}.

We now discuss our results for non-motile limit $(\varepsilon=0)$ of the TSAIM, where the particles of both species only diffuse. A second-order phase transition similar to the one-species AIM is observed between a low-density, high-noise disordered state and a high-density, low-noise ordered state. Fig.~\ref{fig2}(b) shows the mean value of the order parameter $|v_a|$ as a function of the density $\rho_0$ for $\beta_1=0.75$ and several system sizes, which confirms the presence of this order-disorder phase transition. The density $\rho_*$ at which the transition occurs can be determined by the Binder cumulant of the order parameter $U_4 = 1 - \langle v_a^4 \rangle / 3 \langle v_a^2 \rangle^2$ for several system sizes [inset of Fig.~\ref{fig2}(b)]. The Binder cumulant is independent of the system size only at the transition point~\cite{binder1997}, which gives $\rho_*=4.95$ for $\beta_1=0.75$. Repeating this procedure for several temperatures, we compute the temperature-density state diagram for $\varepsilon=0$ as shown in Fig.~\ref{fig2}(c). The transition line can be fitted by the expression $\rho_* = r/(2\beta_1 -1)$, with $r \simeq 2.37$. This expression is compatible with our hydrodynamic theory and is analogous to the one-species AIM.

We now move to the active case ($\varepsilon>0$). Fig.~\ref{fig2}(d) shows the steady-state density profiles $\rho_s(x)$ of a single species in the PF state for varying densities. These profiles are computed in a $1024 \times 128$ domain, integrated along the $y$-direction, and averaged both species and 1000 different times after the steady state is reached. For densities $\rho_0<6.82$, the profiles exhibit phase-separated PF bands similar to Fig.~\ref{fig1}(b), where the density $\rho_s(x)$ takes three different values: maximum (larger than the gas density) in the liquid phase of species $s$ with a positive magnetization $m_s$ due to strong intra-species ferromagnetic interactions; intermediate in the gas phase with zero magnetization; and minimum (smaller than the gas density) in the liquid phase of the other species $-s$ with a negative magnetization $m_s$ due to strong inter-species anti-ferromagnetic interactions. For densities $\rho_0>6.82$, the profiles correspond to an HDPF state as depicted in Fig.~\ref{fig1}(d). From these density profiles, we determine the binodals: $\rho_g = 3.98$ and $\rho_l=6.82$, as the minimum and maximum value of the total density $\rho = \rho_A+\rho_B$, defined such that for $\rho_0<\rho_g$ the system is in the gas phase and $\rho_0>\rho_l$ the system is in a liquid state (either liquid APF or HDPF state, depending on the initial condition). A stable HDPF state emerges solely due to the reciprocal anti-aligning interactions between the two species and therefore, is absent in the one-species flocking models~\cite{AIMprl,AIMpre,APMpre,APMepl,ACMepl,ACMprl}. To the best of our knowledge, this HDPF state has also not been observed in the TSVM~\cite{TSVM} and any other multi-species active matter system.

We subsequently compute the binodals $\rho_g$ and $\rho_l$ for different temperatures and biases, and construct the velocity-density state diagram for $\beta_1=0.75$ [Fig.~\ref{fig2}(e)] and the temperature-density state diagram for $\varepsilon=0.9$ [Fig.~\ref{fig2}(f)], both of which consist of the three typical regions of a flocking system: gas phase ($\rho_0<\rho_g$), liquid state ($\rho_0>\rho_l$), and the liquid-gas coexistence region ($\rho_g<\rho_0<\rho_l$). The three line $\rho_g(T_1)$, $\rho_*(T_1)$ and $\rho_l(T_1)$ merge when $T_1 \to T_c = 2$.

{\bf Hydrodynamic theory of the TSAIM without species flip.} We now present the results obtained from the hydrodynamic theory derived with Eqs.~\eqref{recEq1}-\eqref{recEq4} for $\gamma_2=0$. The non-motile case, for which $\varepsilon=0$ (i.e. $v=0$), exhibits only homogeneous stationary solutions: $\rho(x) = \rho_0$, $m(x)=m_0$, and the order parameters following the equations:
\begin{gather}
v_s = m_0 \tanh \frac{2\beta_1 v_a}{\rho_0} ,  \label{solVS} \\
v_a = \left(\rho_0-\frac{r_1}{2\beta_1} \right) \tanh \frac{2\beta_1 v_a}{\rho_0} \label{solVA}.
\end{gather}
Considering the same average density for each species ($\langle \rho_s(x) \rangle = \rho_0/2$, where $\langle \cdot \rangle$ denotes the spatial averaging here) at initial condition, i.e. $\langle m(x) \rangle = m_0=0$, we get $v_s=0$. From Eq.~\eqref{solVA}, we get a second-order phase transition between a disordered homogeneous solution ($v_a=0$) and an ordered homogeneous solution ($|v_a|>0$). The transition occurs when $2\beta_1-r_1/\rho_0$ is larger than $1$, i.e. for a density larger than $\rho_* = r_1/(2\beta_1 - 1)$ and for a temperature $T_1<T_c = 2$. This expression is compatible with the transition line obtained from the simulations of the microscopic model and shown in Fig.~\ref{fig2}(c).

\begin{figure}[t]
\begin{center}
\includegraphics[width=\columnwidth]{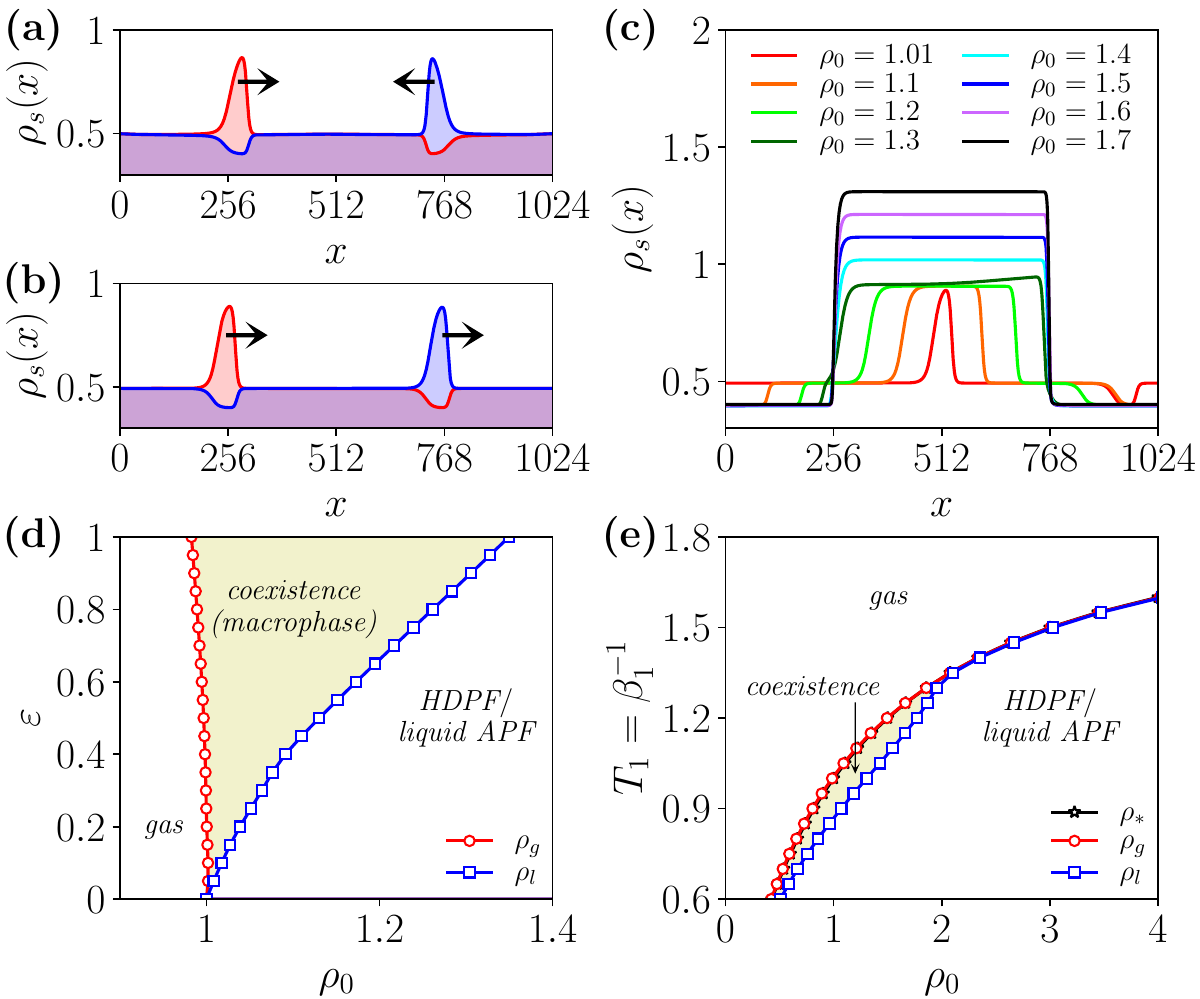}
\caption{{\it Hydrodynamic theory for reciprocal interactions without species flip.} (a--b)~Steady-state density profile (red/blue represent species A/B, respectively) for $\rho_0=1.01$ showing (a)~an APF state and (b)~a PF state. (c)~Steady-state density profile for increasing densities in the PF state. Parameters: $\beta_1=1$, $\varepsilon=0.9$, and $L_x=1024$. (d)~Velocity-density state diagram for $\beta_1=1$. (e)~Temperature-density state diagram for $\varepsilon=0.9$. \label{fig3}}
\end{center}
\end{figure}

Phase-separated density profiles can be observed for a positive velocity ($\varepsilon>0$). Figs.~\ref{fig3}(a) and~\ref{fig3}(b) show steady-state density profiles $\rho_s(x)$ as APF and PF states, respectively. Fig.~\ref{fig3}(c) shows steady-state density profiles $\rho_s$ in the PF state for increasing density. For densities $\rho_0<1.31$, the profiles exhibit phase-separated PF bands similar to Fig.~\ref{fig3}(b), for a right-moving band. The density $\rho_s(x)$ takes three different values: in the liquid phase of species $s$ ($m_s>0$), in the gas phase ($m_s=0$), and in the liquid phase of the other species $-s$ ($m_s<0$). For densities $\rho_0>1.31$, the HDPF state is observed and explained with Eqs.~\eqref{recEq1}-\eqref{recEq4} as follows: in the A-band, the density is homogeneous such that $\rho_A > \rho_B$ and the order parameter $v_a$ obeys the ordered homogeneous solution. Then $v_s$ satisfies Eq.~\eqref{solVS} where $m_0$ is replaced by $\rho_A - \rho_B$, which determines the magnetization of species A: $m_A= (v_s + v_a)/2$ and species B: $m_B= (v_s - v_a)/2$, such that $m_A$ and $m_B$ have different signs. We determine the binodals from these density profiles: $\rho_g =0.986$ and $\rho_l =1.31$. Computing the binodals for several velocities and temperatures, we construct the velocity-density state diagram for $\beta_1=0.75$ [Fig.~\ref{fig3}(d)] and the temperature-density state diagram for $\varepsilon=0.9$ [Fig.~\ref{fig3}(e)]. We recover the three typical regions of a flocking system: gas phase ($\rho_0<\rho_g$), liquid state ($\rho_0>\rho_l$), constituted with the liquid APF state and the HDPF state, and the liquid-gas coexistence region ($\rho_g<\rho_0<\rho_l$).

%%%%%%%%%%%%%%%%%%%%%%%%%%%%%%%%%%
%%% RESULTS: WITH SPECIES FLIP %%%
%%%%%%%%%%%%%%%%%%%%%%%%%%%%%%%%%%

\subsection{Reciprocal interactions with species flip ($\gamma_2 = 0.5$)}

In this subsection, we present the results of the reciprocal TSAIM with species flip, with species interaction given by Eq.~\eqref{wflip2} with $\gamma_2=0.5$, for which the number of particles of each species, $N_A$ and $N_B$, is not conserved. Depending on the initial species magnetization $m_0=(N_A-N_B)/L_xL_y$, the TSAIM dynamics drive the system into different steady states.

\begin{figure*}[t]
\begin{center}
\includegraphics[width=\textwidth]{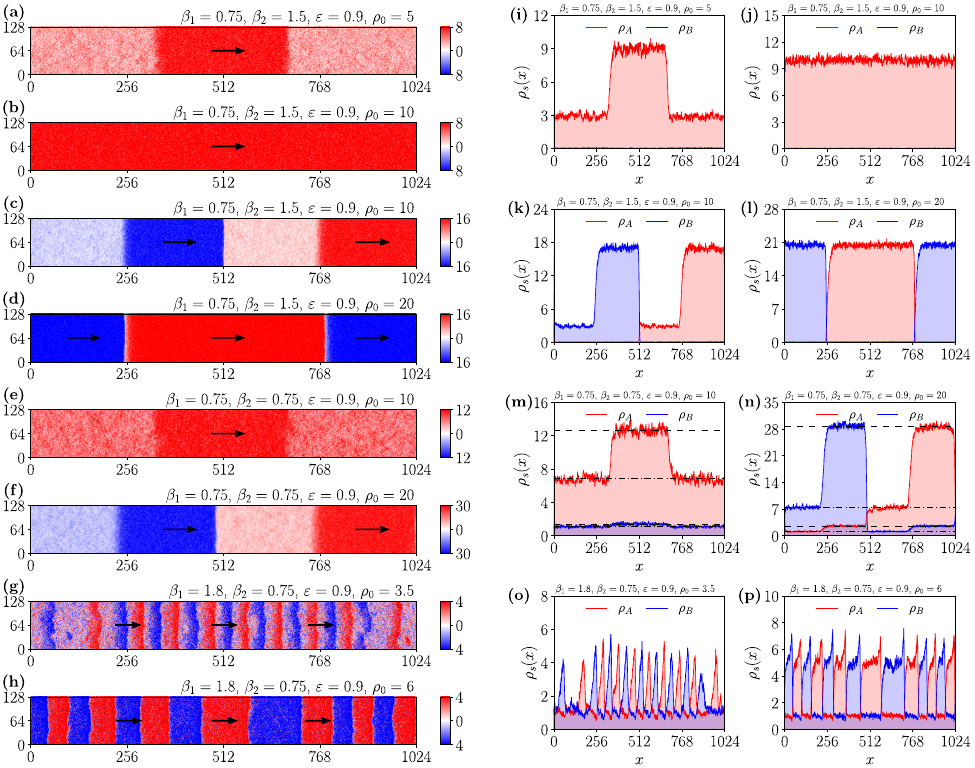}
\caption{{\it Steady states for reciprocal interactions with species flip.} (a--h)~Density snapshots for $\varepsilon=0.9$ in a $1024 \times 128$ domain. (a--d)~Snapshots for $\beta_1=0.75$ and $\beta_2=1.5$: (a)~one-species coexistence state ($\rho_0=5$), (b)~one-species liquid phase ($\rho_0=10$), (c)~two-species coexistence PF state ($\rho_0=10$), and (d)~HDPF state ($\rho_0=20$). (e--f)~Snapshots for $\beta_1=\beta_2=0.75$: (e)~one-species coexistence state ($\rho_0=10$), and (f)~two-species coexistence PF state ($\rho_0=20$). (g--h)~Snapshots for $\beta_1=1.8$ and $\beta_2=0.75$: (g)~two-species coexistence PF state ($\rho_0=3.5$), and (h)~HDPF state ($\rho_0=6$) exhibiting microphase separation. The densities of A and B species are represented with red and blue colors, respectively. A movie (\texttt{movie 5}) of the same can be found at Ref.~\cite{movies}. (i--p)~Density profiles (red/blue represent species A/B, respectively) obtained from the density snapshots shown in (a--h) by integrating along the $y$-axis. The dash-dotted lines represent the densities of the two species in the ferromagnetic gas, while the densities of the two species in the liquid phase are represented by dashed lines, as a guide to the eyes.\label{fig4}}
\end{center}
\end{figure*}

{\bf Steady-state profiles and state diagrams of the TSAIM with species flip.} In this paragraph, we present the results for a hopping rate, given by Eq.~\eqref{eqhop}, with $\theta=0$ and $D=1$. Figs.~\ref{fig4}(a--d) show the steady-state density profiles obtained from the simulation of the microscopic model, for $\beta_1=0.75$ and $\beta_2=1.5$ ($\beta_2 > \beta_1$), starting from a semi-ordered configuration. As shown, the steady state strongly depends on the initial population of each species. At very low densities ($\rho_0<0.8$), the steady state is disordered in both spin and species, forming a {\itshape paramagnetic gas} phase (see  Supplementary Note~4).

Starting from a configuration with all particles of species A ($m_0 = \rho_0$), for $\rho_0 > 0.8$, the system exhibits a liquid-gas phase transition of species A similar to the one-species AIM, where the particles in species B remain disordered for all densities. At low densities ($0.8<\rho_0<2.86$), the particles of species A are disordered in spin, but ordered in species, forming a {\it ferromagnetic gas} phase of species A (see  Supplementary Note~4). At larger densities ($2.86<\rho_0<9.13$), the particles in species A start flocking in a transverse band, showing macrophase-separated density profiles, which characterizes the {\it coexistence} region [Fig.~\ref{fig4}(a)]. At even higher densities ($\rho_0>9.13$), the band of species A spans the whole domain, forming a {\it liquid} phase of species A [Fig.~\ref{fig4}(b)].

But, if we start with an initial configuration where both species are equally populated ($m_0 = 0$), the scenario differs from the one-species AIM and the reciprocal TSAIM without species flip. At low densities ($0.8<\rho_0<2.95$), particles of both species are disordered in spin, but ordered in species, which also leads to a {\it ferromagnetic gas} phase of both species (see  Supplementary Note~4). At higher densities ($2.95 < \rho_0 < 17.0$), both species exhibit transverse band motion, forming macrophase-separated density profiles, with a band of each species followed by its gas phase [Fig.~\ref{fig4}(c)], characterizing the PF {\it coexistence} region. At even higher densities ($\rho_0 > 17.0$), the two bands span the whole domain, with no gas phase present, forming an HDPF state [Fig.~\ref{fig4}(d)], as observed for the reciprocal TSAIM without species flip. However, the liquid APF state is not observed in the presence of species flip. Furthermore, Figs.~\ref{fig4}(b) and~\ref{fig4}(c) are obtained for the same parameters, but different initial configurations, leading to two different steady states: a one-species liquid phase and a two-species phase-separated profile, respectively.

Figs.~\ref{fig4}(e--f) show the steady-state density profiles, for $\beta_1=\beta_2=0.75$, starting from a semi-ordered configuration. For this value of $\beta_2$, the steady states are similar to the ones observed for $\beta_2>1.5$: a paramagnetic gas phase (for $\rho_0<3.95$), a ferromagnetic gas phase (for $3.95<\rho_0<8.0$), a one-species macrophase-separated profile (for $8.0<\rho_0<14.0$) as shown in Fig.~\ref{fig4}(e), a one-species liquid phase (for $\rho_0>14.0$), a two-species macrophase-separated profile (for $8.0<\rho_0<31.0$) as shown in Fig.~\ref{fig4}(d), and an HDPF state (for $\rho_0>31.0$), depending on the initial condition. As shown in Figs.~\ref{fig4}(i--p), the only difference between $\beta_2=1.5$ and $\beta_2=0.75$ lies in the proportion of A and B-particles in each site: for $\beta_2=1.5$, most of the sites of the steady states shown in Figs.~\ref{fig4}(a--d) are occupied by a unique species (either A or B depending on the position), while for $\beta_2=0.75$, all the sites of the steady states shown in Figs.~\ref{fig4}(e--f) are occupied by the two species (but one has a higher population, depending on the position). This behavior is due to the increase of species magnetization $m$, i.e. a stronger order in species, when $\beta_2$ increases.

\begin{figure*}[t]
\begin{center}
\includegraphics[width=\textwidth]{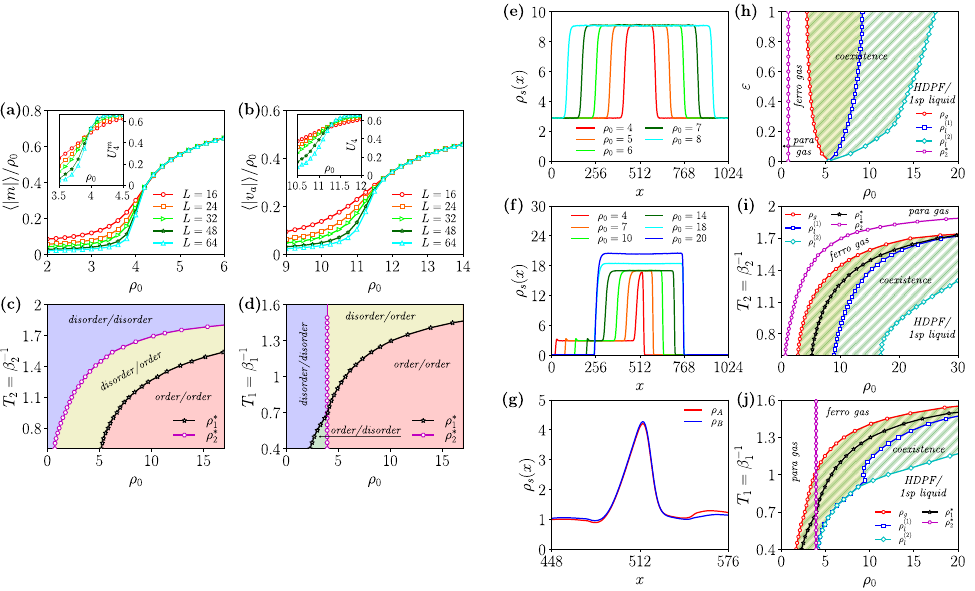}
\caption{{\it Profiles and state diagrams for reciprocal interactions with species flip.} (a--d)~Order parameter and state diagrams for $\varepsilon=0$. (a--b)~Mean value and Binder cumulant (insets) of the order parameters $|m|$ and $|v_a|$ vs $\rho_0$ for $\beta_1=0.75$, $\beta_2=0.75$, and several system sizes. The transitions occur at densities $\rho_2^*=3.95$ and $\rho_1^*=11.25$. (c)~Temperature $T_2$-density state diagram for $\beta_1=0.75$, and (d)~temperature $T_1$-density state diagram for $\beta_2=0.75$. $\rho_0 < \min(\rho_1^*,\rho_2^*)$: spin disordered/species disordered state with $\langle m \rangle=0$ and $\langle v_a \rangle=0$ (blue), $\rho_2^* <\rho_0 < \rho_1^*$: spin disordered/species ordered state with $\langle m \rangle>0$ and $\langle v_a \rangle=0$ (yellow), $\rho_1^* <\rho_0 < \rho_2^*$: spin ordered/species disordered state with $\langle m \rangle = 0$ and $\langle v_a \rangle>0$ (green), and $\rho_0 > \max(\rho_1^*,\rho_2^*)$: spin ordered/species ordered state with $\langle m \rangle>0$ and $\langle v_a \rangle>0$ (red). (e--j)~Profiles and state diagrams for $\varepsilon>0$. (e--f)~Steady-state density profiles (PF state) for $\beta_1=0.75$, $\beta_2=1.5$, and $\varepsilon=0.9$ in a $1024\times 128$ domain, for various densities. The initial configuration is taken with (e)~$m_0=\rho_0$ or (f)~$m_0=0$. (g)~Steady-state density profile of a single band for $\beta_1=1.8$, $\beta_2=0.75$, and $\varepsilon=0.9$ in a $1024\times 128$ domain, corresponding to the microphase separated snapshot shown in Fig.~\ref{fig4}(g). The data are averaged over $1000$ times. (h)~Velocity-density state diagram for $\beta_1=0.75$ and $\beta_2=0.75$. (i)~Temperature $T_2$-density state diagram for $\beta_1=0.75$ and $\varepsilon=0.9$. (j)~Temperature $T_1$-density state diagram for $\beta_2=0.75$ and $\varepsilon=0.9$. \label{fig5}}
\end{center}
\end{figure*}

Figs.~\ref{fig4}(g--h) show the steady-state density profiles, for $\beta_1=1.8$ and $\beta_2=0.75$ ($\beta_1 > \beta_2$), starting from a disordered configuration. Here, the steady state does not depend on the initial population of each species. At low density ($\rho_0<2.11$), the particles are in a disordered state for both spin and species, forming the {\it paramagnetic gas} phase. At higher densities ($2.11< \rho_0<4.67$), both species exhibit transverse band motion, forming {\it microphase-separated} density profiles, with several PF bands of both species [Fig.~\ref{fig4}(g)]. These microphase-separated density profiles have not been observed in the one-species AIM~\cite{AIMprl,AIMpre} or the reciprocal TSAIM without species flip, which only exhibit macrophase-separated profiles. At even higher densities ($\rho_0 > 4.67$), these microphase-separated bands span the whole domain forming an HDPF state [Fig.~\ref{fig4}(h)], similar to Fig.~\ref{fig4}(d) where only two bands are present. This steady state is obtained from a quench and has the signature of a dynamical process with microphase-separated profiles at intermediate times, during the formation of flocks. A steady state with any even number of bands would remain stable within this parameter set. Additional density profiles are shown in Supplementary Note~5. The origin of these microphase-separated profiles is discussed in the context of Fig.~\ref{fig5}.

Figs.~\ref{fig4}(i--p) show the instantaneous steady-state density profiles corresponding to Figs.~\ref{fig4}(a--h). For $\beta_1=0.75$ and $\beta_2=1.5$ ($\beta_2 > \beta_1$), starting from all particles of species A ($m_0=\rho_0$), the density profiles are similar to the one-species AIM: homogeneous gas profiles at low densities, macrophase-separated density profile at intermediate densities [Fig.~\ref{fig4}(i)], and homogeneous liquid density profiles of species A at high densities [Fig.~\ref{fig4}(j)]. Starting from an equal population of both species ($m_0=0$), the macrophase-separated bands of each species are characterized by a PF state, where the high-density liquid band of each species is followed by its own low-density gas phase [Fig.~\ref{fig4}(k)]. At higher $\rho_0$, the low-density gas phases vanish and the high-density liquid bands span the $x$-axis forming an HDPF state [Fig.~\ref{fig4}(l)]. For this value of $\beta_2$, only one of the two species has a positive value (up to fluctuations) at a given position $x$. For $\beta_1=\beta_2=0.75$, starting from all particles of species A, the macrophase-separated density profile exhibits a gaseous background of species B with a slightly higher density inside the liquid band of species A [Fig.~\ref{fig4}(m)]. Starting from an equal population of both species, the macrophase-separated density profiles (within a PF state) exhibit four distinct regions for each species: a maximum density in its own liquid band, a high intermediate density in its own gas phase (located behind its liquid band), a low intermediate density in the liquid band of the other species, and a minimum density in the gas phase of the other species (located behind the liquid band of this species) [Fig.~\ref{fig4}(n)]. Contrary to the reciprocal TSAIM without species flip, the species-species interaction dominates the antiferromagnetic spin-spin interaction meaning that the magnetization of the species $s$ is non-zero only in the liquid band of species $s$, and the density ratio between both species depends only on the species magnetization $m$, such that $\rho_A/\rho_B = (1+M)/(1-M)$, where $M = m/\rho = (\rho_A-\rho_B)/(\rho_A+\rho_B)$ has a constant modulus over $x$ ($M\simeq \pm 0.8$ for $\beta_2=0.75$), leading to the four values taken by $\rho_s$ in gas and liquid regions. For $\beta_1=1.8$ and $\beta_2=0.75$ ($\beta_1 > \beta_2$), multiple alternating high-density bands of species A and B emerge in a PF state on a low-density gas background [Fig.~\ref{fig4}(o)], and as the density increases, the average width and height of these bands expand and span the whole region [Fig.~\ref{fig4}(p)].

Starting from a fully disordered configuration at large $\beta_2$ and $\rho_0$, the system evolves into a steady state where one species completely replaces the other [see Fig.~\ref{fig4}(b)], akin to consensus formation in the binary voter model, where one opinion eventually dominates. A detailed discussion of this evolution is provided in Supplementary Note~6.

We now discuss the non-motile case ($\varepsilon=0$), where the particles of both species only diffuse, in addition to the spin and species interactions. A second-order phase transition occurs for both $m$ and $v_a$ order parameters between a disordered state at low density and high noise, and an ordered state at high density and low noise. Figs.~\ref{fig5}(a--b) show the mean value of the order parameters $|m|$ and $|v_a|$ as a function of the density $\rho_0$ for $\beta_1=\beta_2=0.75$ and several system sizes, which confirms the presence of an order-disorder phase transition for both $m$ and $v_a$. The densities $\rho_2^*$ and $\rho_1^*$ at which both the phase transitions occur can be derived from the Binder cumulant of the two order parameters: $U_4^m = 1 - \langle m^4 \rangle / 3 \langle m^2 \rangle^2$ and $U_4 = 1 - \langle v_a^4 \rangle / 3 \langle v_a^2 \rangle^2$, respectively [insets of Figs.~\ref{fig5}(a--b)]. The Binder cumulant is independent of the system size only at the transition point~\cite{binder1997}, which gives $\rho_2^*=3.95$ for the species phase transition (crossing of $U_4^m$), and $\rho_1^*=11.25$ for the spin-orientation phase transition (crossing of $U_4$) for $\beta_1=\beta_2=0.75$. Repeating this procedure for several $\beta_1$ and $\beta_2$, we compute the temperature-density state diagrams shown in Figs.~\ref{fig5}(c--d). From the flipping rates, given by Eqs.~\eqref{wflip1} and~\eqref{wflip2}, the density $\rho_2^* \equiv \rho_2^*(T_2)$ will only depend on $T_2$ where the density $\rho_1^* \equiv \rho_1^*(T_1,T_2)$ will depend on both the temperatures $T_1$ and $T_2$. The transition line can be fitted by the expression $\rho_2^*(T_2) = r_2/(2\beta_2 -1)$, with $r_2 \simeq 1.85$. This expression is compatible with the hydrodynamic theory and analogous to the one-species AIM. The lines $\rho_1^*$ and $\rho_2^*$ partition the state diagram into four regions in which spin and species can be either disordered or ordered, depending on the values of $\langle v_a \rangle$ and $\langle m \rangle$, respectively. These four non-motile states are similar to those reported for the AT model~\cite{baxter1,baxter2}, for the order parameters $\langle v_s \rangle \sim \langle \sigma \rangle$, $\langle m \rangle \sim \langle s \rangle$, and $\langle v_a \rangle \sim \langle \sigma s \rangle$: (i) a paramagnetic phase where $\langle \sigma \rangle = \langle s \rangle = \langle \sigma s \rangle = 0$ [{\it disorder/disorder} in Figs.~\ref{fig5}(c--d) (blue)], (ii) a symmetric Baxter phase where $\langle \sigma \rangle = \pm \langle s \rangle \ne 0$ and $\langle \sigma s \rangle \ne 0$ [{\it order/order} in Figs.~\ref{fig5}(c--d) (red)], (iii) a $\langle \sigma s \rangle$ phase where $\langle \sigma s \rangle \ne 0$ and $\langle \sigma \rangle = \langle s \rangle = 0$ [{\it order/disorder} in Figs.~\ref{fig5}(c--d) (green)], and (iv) a $\langle s \rangle$ phase where $\langle s \rangle \ne 0$ and $\langle \sigma \rangle = \langle \sigma s \rangle = 0$ [{\it disorder/order} in Figs.~\ref{fig5}(c--d) (yellow)].

We now move to the active case ($\varepsilon>0$). Figs.~\ref{fig5}(e--f) show the steady-state density profiles $\rho_s(x)$ in the PF state obtained from numerical simulations for increasing densities. These profiles are computed in a $1024 \times 128$ domain, integrated along the $y$-direction, and averaged over both species and $1000$ different times after the steady state is reached, following the two different initial populations of each species: $m_0=\rho_0$ in Fig.~\ref{fig5}(e) and $m_0=0$ in Fig.~\ref{fig5}(f). Starting with $m_0=\rho_0$, the density profiles are analogous to the one-species AIM~\cite{AIMprl,AIMpre}. Starting from $m_0=0$ and for densities $\rho_0<17.0$, the profiles exhibit phase-separated PF bands similar to Fig.~\ref{fig4}(c), where the density $\rho_s(x)$ assumes four distinct values corresponding to the liquid phase and the ferromagnetic gas phase, dominated either by species $s$ or by the opposing species $-s$. Starting from $m_0=0$ and for densities $\rho_0>17.0$, the profiles correspond to an HDPF state as depicted in Fig.~\ref{fig4}(d). We determine from these density profiles the value of the binodals for both initial conditions: $\rho_g^{(1)} = 2.86$ and $\rho_l^{(1)}=9.13$, as the minimum and maximum value of $\rho_A$ when only species A is present, and $\rho_g^{(2)} = 2.95$ and $\rho_l^{(2)}=17.0$, as the minimum and maximum value of the total density $\rho = \rho_A+\rho_B$ when both species are present. As we can see $\rho_g^{(1)} \simeq \rho_g^{(2)} \equiv \rho_g$ since the system is in the same gas phase for both scenarios when $\rho_0 < \rho_g$. However, we can see that $\rho_l^{(1)} < \rho_l^{(2)}$ since the two liquid states are completely different and the antiferromagnetic interaction between the two species makes the value of the binodal higher.

Fig.~\ref{fig5}(g) shows the steady-state density profile $\rho_s(x)$ of a single band obtained from numerical simulations, corresponding to the microphase separated snapshot shown in Fig.~\ref{fig4}(g). This profile is computed in a $1024 \times 128$ domain by first performing a running average on the time-dependent density profiles, integrated along the $y$-direction, to merge the multiple bands corresponding to the microphase separation. The resulting profiles are then averaged over 1000 times, and both species. We determine the gas and liquid binodals: $\rho_g = 2.11$ and $\rho_l^{(2)} = 4.67$, respectively, from this density profile.

We compute analogously the binodals $\rho_g$, $\rho_l^{(1)}$, and $\rho_l^{(2)}$ for different temperatures $T_1$, $T_2$, and biases $\varepsilon$, and construct the velocity-density state diagram for $\beta_1=0.75$ and $\beta_2=1.5$ [Fig.~\ref{fig5}(h)], the temperature $T_2$-density state diagram for $\beta_1=0.75$ and $\varepsilon=0.9$ [Fig.~\ref{fig5}(i)], and the temperature $T_1$-density state diagram for $\beta_2=0.75$ and $\varepsilon=0.9$ [Fig.~\ref{fig5}(j)], which consists of four regions: paramagnetic gas phase ($\rho_0<\rho_2^*$), ferromagnetic gas phase ($\rho_2^*<\rho_0<\rho_g$), coexistence region ($\rho_g<\rho_0<\rho_l$), and liquid state ($\rho_0>\rho_l$), where $\rho_l$ can either be $\rho_l^{(1)}$ or $\rho_l^{(2)}$, depending on the initial configuration. This liquid state is characterized by a one-species liquid phase ({\it 1sp liquid}) for $\rho_0>\rho_l^{(1)}$, as shown in Fig.~\ref{fig4}(b), or by an HDPF state for $\rho_0>\rho_l^{(2)}$, as depicted in Fig.~\ref{fig4}(d). The coexistence region can be further decomposed into three subregions: (i) $\rho_2^*<\rho_g<\rho_1^*$ where a ferromagnetic gas phase coexists with a macrophase-separated liquid band [Fig.~\ref{fig4}(a)], (ii) $\rho_g<\rho_2^*<\rho_1^*$ where a paramagnetic gas phase coexists with a macrophase-separated liquid band [Fig.~\ref{fig4}(c)], and (iii) $\rho_g<\rho_1^*<\rho_2^*$ where a paramagnetic gas phase coexists with microphase-separated liquid bands emerging from the frustration of the spin ordered but species disordered state [Fig.~\ref{fig4}(g)]. Note that the liquid binodals, $\rho_l^{(1)}$ and $\rho_l^{(2)}$, will always be larger than $\rho_g$, $\rho_1^*$, and $\rho_2^*$ since flocking is possible only when both species and spin are ordered.

%Similarly to the TSAIM without species flip, the liquid states are metastable when the diffusion is small compared to the velocity, as shown in appendix~\ref{sec5b}. The steady-state consists of left- and right-flocking clusters arranged in alternating stripes of A and B species along the y-axis. These stripes exhibit long-range order in species within each stripe but display short-range order in spin, akin to the behavior observed in the one-species AIM. Additionally, the TSAIM with species flip also demonstrates MIIP (see appendix~\ref{sec5b}).

\begin{figure}[t]
\begin{center}
\includegraphics[width=\columnwidth]{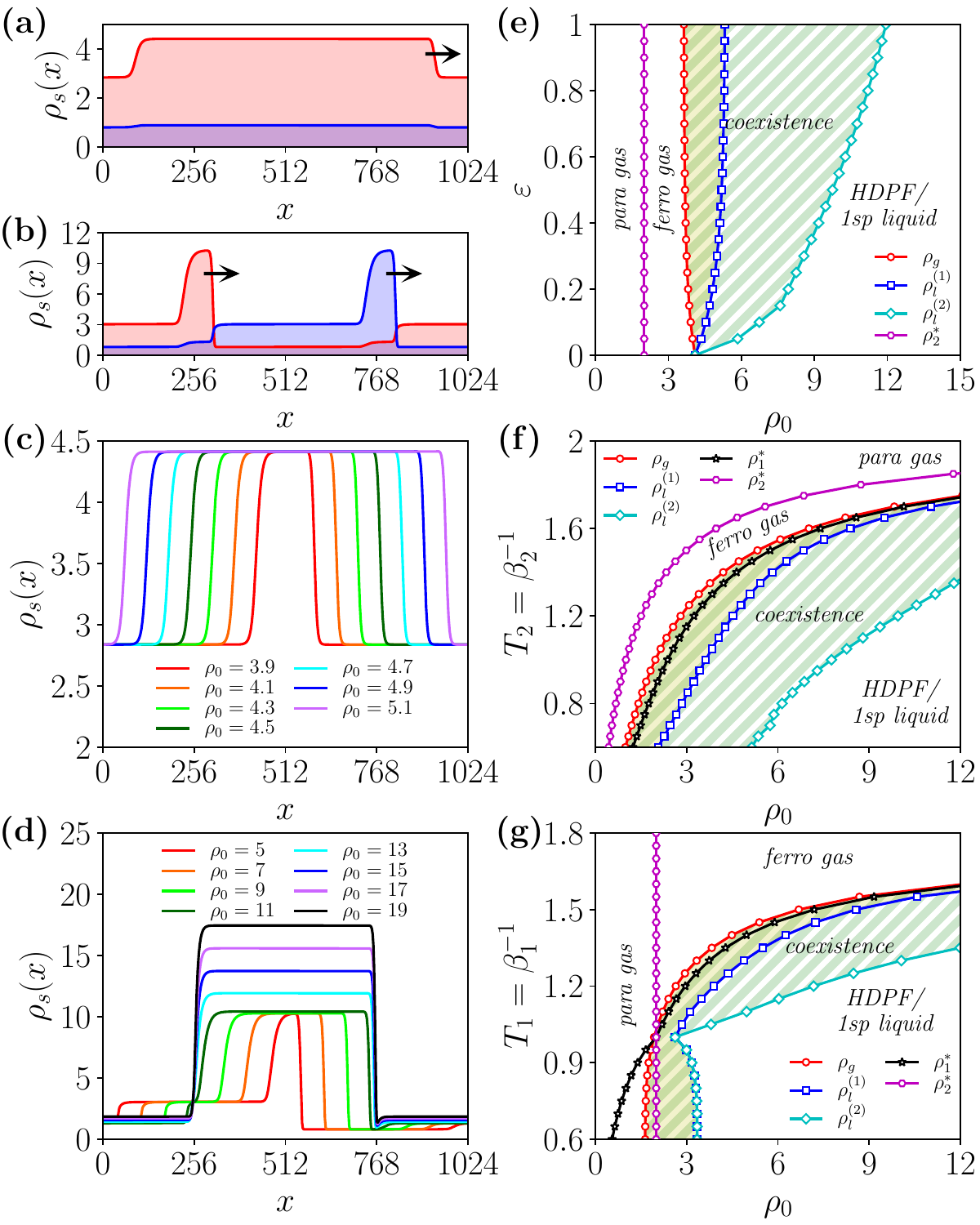}
\caption{{\it Hydrodynamic theory for reciprocal interactions with species flip.} (a--b)~Steady state profiles (red/blue represent species A/B, respectively) for same density $\rho_0=5$ but different initial species magnetization: (a)~$m_0=5$, and (b)~$m_0=0$. (c--d)~Steady-state density profiles (in PF state) for different densities and initial species magnetization: (c)~$m_0=\rho_0$, and (d)~$m_0=0$. Parameters: $\beta_1=0.75$, $\beta_2=0.75$, $\varepsilon=0.9$, and $L_x=1024$. (e)~Velocity-density state diagram for $\beta_1=\beta_2=0.75$. (f)~Temperature $T_2$-density state diagram for $\beta_1=0.75$ and $\varepsilon=0.9$. (g)~Temperature $T_1$-density state diagram for $\beta_2=0.75$ and $\varepsilon=0.9$. \label{fig6}}
\end{center}
\end{figure}

{\bf Hydrodynamic theory of the TSAIM with species flip.} We now present the results obtained from the hydrodynamic theory derived with Eqs.~\eqref{recEq1}-\eqref{recEq4} with $\gamma_2 \ne 0$. Here, we will restrain to the special case $r_1=r_2=r$, leading to ${\overline \gamma_2} / {\overline \gamma_1} = \gamma_2/\gamma_1 \equiv \gamma$. The non-motile case, for which $\varepsilon=0$ (i.e. $v=0$), exhibits only homogeneous stationary solutions: $\rho(x) = \rho_0$, and
\begin{gather}
m = \left(\rho_0- \frac{r}{2\beta_2} \right) \tanh \frac{2\beta_2 m}{\rho_0} , \label{SolM}\\
v_s = m \tanh \frac{2\beta_1 v_a}{\rho_0}, \label{SolVS2}\\
\left(\rho_0- \frac{r}{2\beta_1} \right) \sinh \frac{2\beta_1 v_a}{\rho_0} - v_a \cosh \frac{2\beta_1 v_a}{\rho_0} \nonumber \\ + \gamma \left[ v_s \sinh\frac{2\beta_2 m}{\rho_0} -  v_a \cosh \frac{2\beta_2 m}{\rho_0} \right] =0. \label{SolVA2}
\end{gather}
From Eq.~\eqref{SolM}, we get a second-order phase transition between a disordered homogeneous solution where $m=0$ and an ordered homogeneous solution $|m|>0$. The transition occurs when $2\beta_2-r/\rho_0$  is larger than 1, i.e. for a density larger than $\rho_2^* = r/(2\beta_2 - 1)$ and for a temperature $T_2<T_c = 2$. This expression is compatible with the transition line obtained from simulations of the microscopic model and shown in Fig.~\ref{fig5}(c). From Eqs.~\eqref{SolVS2}-\eqref{SolVA2}, we get a second-order phase transition between a disordered homogeneous solution where $v_a=0$ and an ordered homogeneous solution $|v_a|>0$. The transition occurs for a density larger than $\rho_1^*$, solution of the equation:
\begin{equation}
\rho_1^* = \frac{r}{2\beta_1 -1 + \gamma[(\beta_1/\beta_2) M_* \sinh M_* - \cosh M_*]},
\end{equation}
with $M_* = 2 \beta_2 m_* / \rho_1^*$ where $m_*$ is a stable solution of Eq.~\eqref{SolM} with $\rho_0=\rho_1^*$, i.e. $m_*=0$ for $\rho_1^*<\rho_2^*$ and $m_*>0$ for $\rho_1^*>\rho_2^*$. The solution is trivial for $\rho_1^*<\rho_2^*$: $\rho_1^* = r/(2\beta_1 -1- \gamma)$, but not analytically tractable for $\rho_1^*>\rho_2^*$. Numerical solutions tell us that $\rho_1^*$ is continuous as function of $T_1$ and $T_2$, but its derivative exhibits a discontinuity at the temperature $T_1$ such that $\rho_1^*(T_1,T_2) = \rho_2^*(T_2)$, i.e. $T_1 =  T_2 /(1+\gamma T_2/2)$.

Phase-separated density profiles can be observed for a positive velocity ($\varepsilon>0$). Figs.~\ref{fig6}(a--d) show the steady-state density profiles $\rho_s(x)$ in the PF state. Two different initial average species magnetization are considered: $\langle m(x) \rangle = m_0 = \rho_0$ for Figs.~\ref{fig6}(a,c) and $\langle m(x) \rangle = m_0 = 0$ for Figs.~\ref{fig6}(b,d). Analogous to the results obtained from the simulations of the microscopic model, two kinds of phase-separated density profiles are observed for the same density $\rho_0=5$: one-species flocking when starting with $m_0=\rho_0$ [Fig.~\ref{fig6}(a)], akin to the one-species AIM~\cite{AIMprl,AIMpre}, and two-species flocking when starting with $m_0=0$ [Fig.~\ref{fig6}(b)], where the density $\rho_s(x)$ takes four different values for the same reasons discussed in Fig.~\ref{fig4}(n). Figs.~\ref{fig6}(c--d) show theses steady-state density profiles $\rho_s(x)$ with increasing densities. Starting with $m_0=\rho_0$, the density profiles are analogous to the one-species AIM~\cite{AIMprl,AIMpre} [Fig.~\ref{fig6}(c)], and starting with $m_0=0$, the profiles exhibit phase-separated PF bands similar to Fig.~\ref{fig6}(b) for densities $\rho_0<11.59$ and a HDPF state for densities $\rho_0>11.59$ [Fig.~\ref{fig6}(d)].

We determine the binodals from the minimum and maximum density $\rho= \rho_A + \rho_B$ of these phase-separated profiles: $\rho_g^{(1)}=3.64$ and  $\rho_l^{(1)}=5.30$ when one species is present, and $\rho_g^{(2)}=3.60$ and  $\rho_l^{(2)}=11.59$ when two species are present in the steady state. As for simulations of the microscopic model, we have $\rho_g^{(1)} \sim \rho_g^{(2)} \equiv \rho_g$ and $\rho_l^{(1)} < \rho_l^{(2)}$. Computing the binodals for several velocities $\varepsilon$ and temperatures $T_1$ and $T_2$, we construct the velocity-density state diagram for $\beta_1=\beta_2=0.75$ [Fig.~\ref{fig6}(e)], the temperature $T_2$-density state diagram for $\varepsilon=0.9$ and $\beta_1=0.75$ [Fig.~\ref{fig6}(f)], and the temperature $T_1$-density state diagram for $\varepsilon=0.9$ and $\beta_2=0.75$ [Fig.~\ref{fig6}(g)]. We recover the four regions observed with simulations of the microscopic model: paramagnetic gas phase ($\rho_0<\rho_2^*$), ferromagnetic gas phase ($\rho_2^*<\rho_0<\rho_g$), coexistence region ($\rho_g<\rho_0<\rho_l$), and liquid state ($\rho_0>\rho_l$), where $\rho_l$ can either be $\rho_l^{(1)}$ or $\rho_l^{(2)}$, depending on the initial configuration. This liquid state is characterized by a one-species liquid phase ({\it 1sp liquid}) for $\rho_0>\rho_l^{(1)}$, or by an HDPF state for $\rho_0>\rho_l^{(2)}$. However, the coexistence region will be slightly different here: the gas binodal $\rho_g$ will always be between $\rho_1^*$ and $\rho_2^*$ leading to two subregions: (i) $\rho_2*<\rho_g<\rho_1^*$ where a ferromagnetic gas phase ($m>0$ and $v_a=0$) coexists with a liquid band ($m>0$ and $v_a>0$), and (ii) $\rho_1*<\rho_g<\rho_2^*$ where a APF state ($m=0$ and $v_a>0$) coexists with a liquid band ($m>0$ and $v_a>0$). This second subregion is unstable in the presence of fluctuations, leading to the microphase-separated profiles observed in Figs~\ref{fig4}(g--h).

%%%%%%%%%%%%%%%%%%%%%%%%%%%%%%%%%%%%
%%% RESULTS: WITH NR INTERACTION %%%
%%%%%%%%%%%%%%%%%%%%%%%%%%%%%%%%%%%%

\subsection{Non-reciprocal interactions}

In this subsection, we consider the TSAIM with non-reciprocal interactions (NRTSAIM) which is a minimal microscopic model of non-reciprocal flocking with two species having competing interests: species A aligns with species B with an interaction strength $J_{AB}$ $(0 \leq J_{AB} \leq J_1)$ while species B anti-aligns with species A with interaction strength $J_{BA}$ $(-J_1 \leq J_{BA} \leq 0)$. We further define the dimensionless variables ${\cal J}_{ss'} = J_{ss'}/J_1$ and $\beta_1=\beta J_1$. Here, we will again consider an equal population of both species, $N_A=N_B=N/2$, i.e. $m_0=0$.

For the results presented in this subsection, we initialize the system in either an ordered or semi-ordered configuration. In an ordered configuration, particles of both species possess the same spin $\sigma$, whereas in the semi-ordered configuration, high-density bands of both species are arranged in either a PF or an APF state. If simulations start from a disordered initial configuration in the presence of non-reciprocal frustration, the system tends to remain disordered. This is due to the continuous switching of the spin $\sigma$ of a particle between its two states $(\pm 1)$, resulting in a disordered steady state.

{\bf Steady-state profiles and state diagrams of the NRTSAIM.} In this paragraph, we present the results for a hopping rate, given by Eq.~\eqref{eqhop}, with $\theta=0$ and $D=1$. First, we consider the non-reciprocal interaction as ${\cal J}_{AB} = - {\cal J}_{BA} \equiv {\cal J}_{\rm NR}$, where the coupling constant of A-particles with B-particles, ${\cal J}_{AB}>0$, is exactly the opposite of the coupling constant of B-particles with A-particles, ${\cal J}_{BA}<0$. 

\begin{figure*}[t]
\begin{center}
\includegraphics[width=\textwidth]{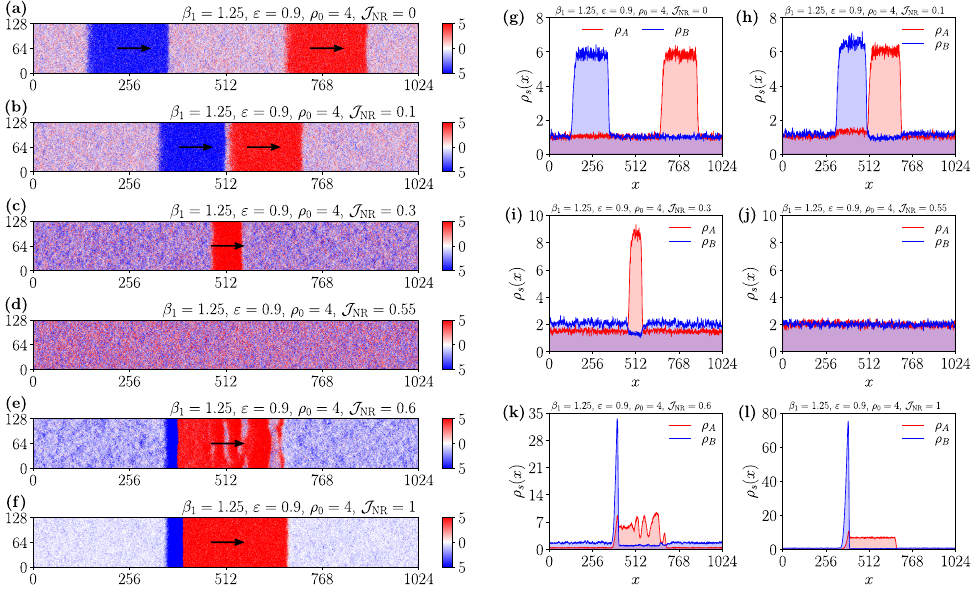}
\caption{{\it Steady states for non-reciprocal interactions.} (a--f)~Density snapshots for $\beta_1=1.25$, $\varepsilon=0.9$, $\rho_0=4$, and increasing non-reciprocal interaction: (a)~${\cal J}_{\rm NR}=0$, (b)~${\cal J}_{\rm NR}=0.1$, (c)~${\cal J}_{\rm NR}=0.3$, (d)~${\cal J}_{\rm NR}=0.55$, (e)~${\cal J}_{\rm NR}=0.6$, and (f)~${\cal J}_{\rm NR}=1$ in a $1024 \times 128$ domain. The densities of
A and B species are represented with red and blue colors, respectively. A movie (\texttt{movie 8}) of the same can be found at Ref.~\cite{movies}. (g--l)~Density profiles (red/blue represent species A/B, respectively) for increasing ${\cal J}_{\rm NR}$, obtained from the density snapshots shown in (a--f) by integrating along the $y$-axis.}
\label{fig7}
\end{center}
\end{figure*}

We first analyze the effect of self-propulsion ($\varepsilon>0$). Starting from the same initial configuration, Figs.~\ref{fig7}(a--f) show all the possible steady-state density configurations exhibited by the NRTSAIM for large self-propulsion $\varepsilon=0.9$ as the strength of the non-reciprocal interaction ${\cal J}_{\rm NR}$ is increased. Fig.~\ref{fig7}(a) and Fig.~\ref{fig7}(b) exhibit a flocking state of two species (in a PF state) for ${\cal J}_{\rm NR}=0$ and ${\cal J}_{\rm NR}=0.1$, respectively. In Fig.~\ref{fig7}(b), the B-band (blue) appears narrower and denser compared to the A-band whereas in Fig.~\ref{fig7}(a), both bands exhibit equal width and density due to the absence of inter-species interactions. This occurs because the slight increase in the non-reciprocity increases the velocity of the B-band and propels it near to the opposite extremity of the A-band (relative to the band propulsion direction) as B-particles respond to the emerging non-reciprocal environment. Despite the growing strength of non-reciprocal interactions, the two-species flocking behavior persists up to a point, after which the system transitioned to a flocking state of only species A, as shown in Fig.~\ref{fig7}(c), for ${\cal J}_{\rm NR}=0.3$. However, as we increase ${\cal J}_{\rm NR}$ further, a transition to a fully disordered state occurs in both species [Fig.~\ref{fig7}(d)], followed by run-and-chase dynamics at higher values of ${\cal J}_{\rm NR}$ [Figs.~\ref{fig7}(e--f)]. This run-and-chase dynamics emerges from the nucleation of an A-band in the gas phase in front of the B-band, since A-particles tend to align with B-particles. The B-particles tend to avoid them because of the non-reciprocal interaction, resulting in a substantial accumulation of B-particles, which leads to a denser B-band, and a slowing down of the B-band velocity. Therefore, the arrangement shown in Fig.~\ref{fig7}(f) exemplifies a highly efficient non-reciprocal configuration, facilitating the B-particles to maintain the {\it maximum distance} from the chasing A-particles. The velocity $c \sim 1.96$ of the flocking bands [Figs.~\ref{fig7}(a--b)] is larger than the self-propulsion velocity of the particles $v=2D\varepsilon = 1.8$ for small non-reciprocity, as also observed in the one-species AIM~\cite{AIMprl,AIMpre}. However, in the run-and-chase state [Figs.~\ref{fig7}(e--f)], the velocity $c \sim 1.64$ of the B-band is smaller than $v$ since the non-reciprocal interaction slows down the dynamics of the B-band.

Fig.~\ref{fig7} qualitatively evaluates the NRTSAIM states via steady-state snapshots as the non-reciprocal interaction strength increases. This qualitative assessment is complemented by the corresponding density profiles shown in Figs.~\ref{fig7}(g--l), which provide a more quantitative picture of the phenomenon. We observe two identical yet distinct density profiles of species A and B in Fig.~\ref{fig7}(g) as the inter-species interaction is zero. The profiles signify that the system is in the liquid-gas coexistence regime, where each species exhibits order solely within its respective band while remaining disordered elsewhere. For ${\cal J}_{\rm NR}=0.1$ [Fig.~\ref{fig7}(h)], with the emergence of non-reciprocity, the liquid density within the B-profile surpasses that of the corresponding A-profile, as B-particles try to move away from the high-density A-band. However, within the B-band, A-particles align with B-particles, both sharing the same spin state $\sigma$ and moving in the same direction. In contrast, within the A-band, B-particles are anti-aligned with A-particles, moving in the opposite direction with an opposite spin state. A similar effect occurs in the one-species flocking band of species A [Fig.~\ref{fig7}(i) for ${\cal J}_{\rm NR}=0.3$], where non-reciprocity causes contrasting behaviors in B-particles: they are in a disordered state outside the A-band, but in an anti-aligned ordered state inside the band. With a continued increase in non-reciprocal frustration, both species transition into a disordered state [Fig.~\ref{fig7}(j)]. Subsequently, a run-and-chase state emerges for large ${\cal J}_{\rm NR}$ [Figs.~\ref{fig7}(k--l)] where the bands of each species are coupled. In this run-and-chase state, the maximum density of the B-profile increases with ${\cal J}_{\rm NR}$ as an increasing number of B-particles try to escape contact with the moving A-band.

\begin{figure*}[t]
\begin{center}
\includegraphics[width=\textwidth]{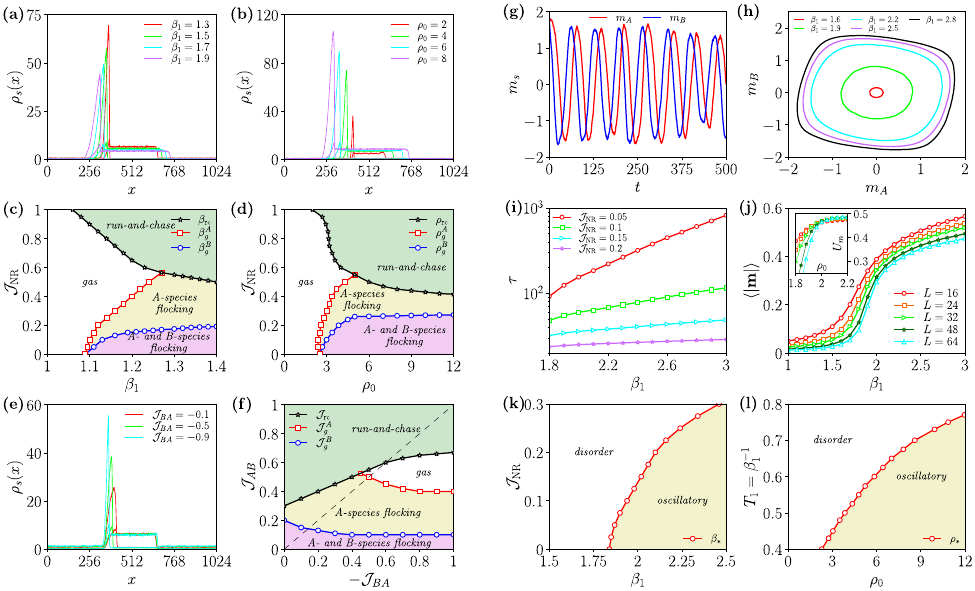}
\caption{{\it Profiles and state diagrams for non-reciprocal interactions.} (a--b) Steady-state density profiles of species A (flat profiles) and species B (sharp peaks), shown for (a)~different $\beta_1$ with fixed $\rho_0=4$, and (b)~varying $\rho_0$ with fixed $\beta_1=1.25$. Parameters: $\varepsilon=0.9$, ${\cal J}_{\rm NR}=1$, $L_x=1024$, $L_y = 128$. (c)~(${\cal J}_{\rm NR}$,$\beta_1$) state diagram for $\rho_0=4$ and $\varepsilon=0.9$; and (d)~(${\cal J}_{\rm NR}$,$\rho_0$) state diagram for $\beta_1=1.25$ and $\varepsilon=0.9$. (e)~Steady-state density profiles for different values of ${\cal J}_{BA}$ and fixed ${\cal J}_{AB}=0.7$, $\beta_1=1.25$, $\varepsilon=0.9$, $\rho_0=4$ in a $1024 \times 128$ domain. (f)~(${\cal J}_{AB}$,$-{\cal J}_{BA}$) state diagram for $\beta_1=1.25$, $\varepsilon=0.9$, and $\rho_0=4$. The dotted line represents ${\cal J}_{AB} = - {\cal J}_{BA}$ as a guide to the eyes. (g--l)~Non-motile NRTSAIM ($\varepsilon=0$): (g)~Time-evolution of $m_A$ and $m_B$ exhibiting an oscillatory state for $\beta_1=2.2$, $\rho_0=4$, and ${\cal J}_{\rm NR}=0.1$ in a $64\times64$ domain. (h)~Phase portrait for $\rho_0=4$, ${\cal J}_{\rm NR}=0.1$, and increasing $\beta_1$ in a $64 \times 64$ domain. (i)~Oscillation period vs $\beta_1$ for several ${\cal J}_{\rm NR}$ couplings and $\rho_0=4$. (j)~Mean value and Binder cumulant (inset) of the vector order parameter ${\bf m}=(m_A,m_B)$ vs $\beta_1$ for $\rho_0=4$, ${\cal J}_{\rm NR}=0.1$, and several system sizes. The transition occurs at $\beta_*=1.94$. (k)~$({\cal J}_{\rm NR},\beta_1)$ state diagram for $\rho_0=4$, and (l)~$(T_1,\rho_0)$ state diagram for ${\cal J}_{\rm NR}=0.1$.}
\label{fig8}
\end{center}
\end{figure*}

To illustrate the influence of density and noise on the run-and-chase state, we present the steady-state density profiles of species A and B for ${\cal J}_{\rm NR}=1$ in Figs.~\ref{fig8}(a--b). Regardless of whether system density increases at fixed noise or $\beta_1$ increases at fixed density, the system remains in the run-and-chase state, without transitioning to an ordered phase. As $\beta_1$ is increased, with fixed density $\rho_0$, the run-and-chase state exhibits wider A and B-bands and a decrease in the maximal density of each band, due to stronger intra-species interactions [Fig.~\ref{fig8}(a)]. As the total density is increased, with fixed $\beta_1$, the run-and-chase state shows broader but denser bands [Fig.~\ref{fig8}(b)]. A higher density leads more B-particles to move within the narrow, high-density B-band to avoid the expanding A-band. However, since A-particles are aligned with B-particles, the density of A-particles noticeably increases within the high-density B-band.

To comprehensively illustrate the influence of noise and density on the NRTSAIM steady states, we construct the (${\cal J}_{\rm NR}$,$\beta_1$) and (${\cal J}_{\rm NR},\rho_0$) state diagrams, depicted in Figs.~\ref{fig8}(c--d). These state diagrams exhibit four regions representing the following states: A- and B-species flocking, A-species flocking, disordered gas, and run-and-chase states. When ${\cal J}_{\rm NR}=0$, denoting the absence of inter-species interaction, we observe the emergence of flocking bands of both species, as noise decreases or density increases. As ${\cal J}_{\rm NR}$ is increased for $\beta_1=1.25$ and $\rho_0 = 4$, the system goes through all the possible states of the NRTSAIM, as shown in Fig.~\ref{fig7}. As $\beta_1$ and $\rho_0$ are increased, the transition from a flocking state to the disordered state occurs at larger values of ${\cal J}_{\rm NR}$, while the transition to the run-and-chase state appears at smaller values of ${\cal J}_{\rm NR}$, as this state becomes more favorable even with weaker non-reciprocal coupling.

We now shift our focus to the scenario where the magnitudes of the inter-species coupling strengths may differ from each other (${\cal J}_{AB} \ne - {\cal J}_{BA}$, but keeping ${\cal J}_{AB} \ge 0$ and ${\cal J}_{BA} \le 0$). We observe that the impact of ${\cal J}_{AB}$ stands in stark contrast to that of ${\cal J}_{BA}$. By maintaining a constant ${\cal J}_{AB}=0.7$, the system exhibits a run-and-chase state for any values of ${\cal J}_{BA}$. An increase of $|{\cal J}_{BA}|$ further consolidates the run-and-chase state, as evidenced by the density profiles of the B-species [Fig.~\ref{fig8}(e)].

Subsequently, we construct a $({\cal J}_{AB},-{\cal J}_{BA})$ state diagram in Fig.~\ref{fig8}(f) to illustrate the comparative influence of ${\cal J}_{AB}$ and ${\cal J}_{BA}$ in defining the steady-states of the NRTSAIM. To illustrate the effect of nonreciprocity, let us move horizontally in the state diagram keeping ${\cal J}_{AB}$ constant, from ${\cal J}_{BA}=0$ to ${\cal J}_{BA}=-1$ (increasing the anti-alignment interaction). For small ${\cal J}_{AB}$, the system exhibits a two-species or one-species flocking state, illustrated in Figs.~\ref{fig7}(a--c). For ${\cal J}_{AB} \ge 0.7$, regardless of the magnitude of the coupling ${\cal J}_{BA}$, the system consistently manifests a run-and-chase state, as shown in Figs.~\ref{fig7}(e--f), where the velocity of the B-band slowly decreases with $-{\cal J}_{BA}$.  On the other hand, if we now move vertically in the state diagram keeping ${\cal J}_{BA}$ constant, the system transitions from a two-species flocking state to a run-and-chase state, via a one-species flocking state, when ${\cal J}_{AB}$ is increased. For large ${\cal J}_{BA}$, this transition occurs through an intermediate disordered gas state. Moving along the diagonal dotted line ${\cal J}_{AB} = - {\cal J}_{BA}$, from the bottom-left to the upper-right corners, we go through the steady states as presented in Fig.~\ref{fig7}. Hence, it appears that ${\cal J}_{AB}$ assumes a more critical role in determining the emergence of the steady states than ${\cal J}_{BA}$. To emphasize, a strong preference of species A to B propels the system toward the run-and-chase state, even for a slight aversion of B towards A.

{\bf Non-motile NRTSAIM ($\varepsilon=0$).} We now examine the scenario of non-motile NRTSAIM ($\varepsilon=0$), where particles of both species only diffuse. In the absence of non-reciprocity $({\cal J}_{\rm NR}=0)$, a phase transition occurs from a disordered state at low density and large noise to an ordered state at large density and low noise. With the introduction of non-reciprocity via ${\cal J}_{\rm NR}$, we witness the emergence of a dynamical oscillatory state (or swap state)~\cite{vitteliNRAIM}, characterized by oscillations in the magnetizations of both species over time [Fig.~\ref{fig8}(g)], where $m_A$ is in late quadrature with $m_B$. The magnitudes of these oscillations fluctuate over time, and generally increase with $\beta_1$ and decrease with ${\cal J}_{\rm NR}$. Fig.~\ref{fig8}(h) shows the phase portraits ($m_A$,$m_B$) and illustrates the transition from the disordered state to the oscillatory state. In the disordered state, $m_A=m_B=0$ is the only stable fixed point, while the oscillatory state is depicted by stable limit cycles, with their area increasing as $\beta_1$ increases. Furthermore, the time period of these oscillations $\tau$ increases exponentially with $\beta_1$, and decreases with ${\cal J}_{\rm NR}$ [Fig.~\ref{fig8}(i)]. Indeed, a longer period (signifying stronger ordering) arises from weaker noise (larger $\beta_1$) and low non-reciprocal strength.

To characterize the transition points between the disordered state and the oscillatory state, we define a vector order parameter ${\bf m}=(m_A,m_B)$. Fig.~\ref{fig8}(j) shows the mean value of this order parameter $|{\bf m}|$ as a function of $\beta_1$ for several system sizes, characteristic of a transition between a disordered state and a quasi-ordered state~\cite{baek2009}. The inverse temperature $\beta_*$ at which this disordered/oscillatory transition occurs is determined by the Binder cumulant of the vector order parameter $U_m = 1 - \langle {\bf m}^4 \rangle / 2 \langle {\bf m}^2 \rangle^2$, with ${\bf m}^2  = m_A^2 + m_B^2$, for several system sizes [inset of Fig.~\ref{fig8}(j)]. The Binder cumulant is independent of the system size only at the transition point~\cite{binder1997}, which gives $\beta_*=1.94$ for $\rho_0=4$ and ${\cal J}_{\rm NR}=0.1$. By repeating this procedure for varying $\rho_0$ and ${\cal J}_{\rm NR}$, we compute the (${\cal J}_{\rm NR}$,$\beta_1$) and the ($T_1$,$\rho_0$) state diagrams, shown in Figs.~\ref{fig8}(k--l), respectively, displaying two steady-states: the disordered and the oscillatory states. However, for ${\cal J}_{\rm NR} = 0$ in Fig.~\ref{fig8}(k), the transition occurs between the disordered and the ordered states, where the ordered state can be interpreted as an oscillatory state with an infinite oscillation period. For any nonzero ${\cal J}_{\rm NR}$, regardless of how small, the oscillation period does not diverge at a finite $\beta_1$ value, showing that no ordered state can be reached [Fig.~\ref{fig8}(i)].

\begin{figure*}[t]
\begin{center}
\includegraphics[width=\textwidth]{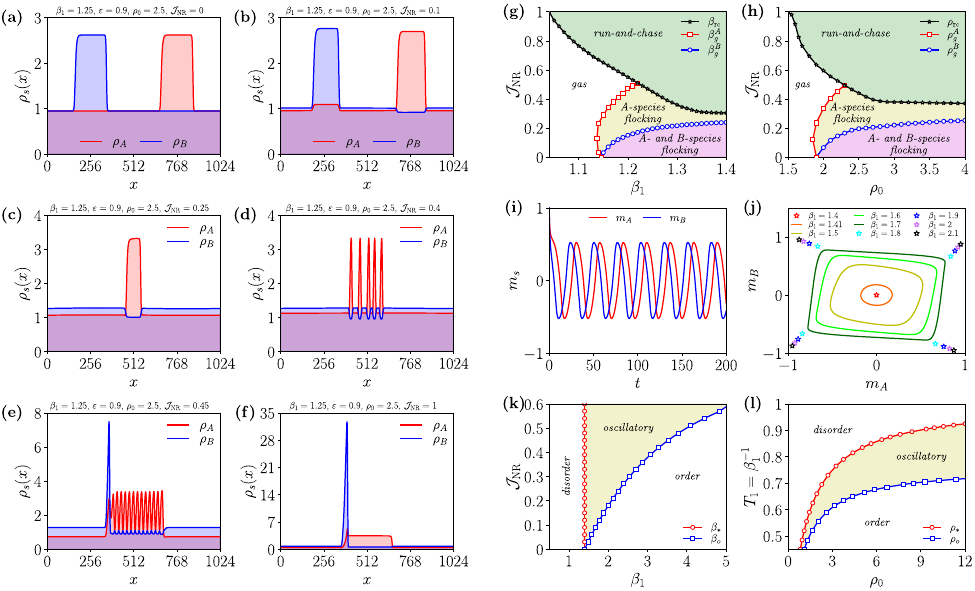}
\caption{{\it Hydrodynamic theory for non-reciprocal interactions.} (a--f)~Steady-state density profiles (red/blue represent species A/B, respectively) for $\beta_1=1.25$, $\varepsilon=0.9$, $\rho_0=2.5$, $L_x=1024$, and increasing ${\cal J}_{\rm NR}$: (a)~${\cal J}_{\rm NR} = 0$, (b)~${\cal J}_{\rm NR} = 0.1$, (c)~${\cal J}_{\rm NR} = 0.25$, (d)~${\cal J}_{\rm NR} = 0.4$, (e)~${\cal J}_{\rm NR} = 0.45$, and (f)~${\cal J}_{\rm NR} = 1$. (g)~$({\cal J}_{\rm NR},\beta_1)$ state diagram for $\rho_0=2.5$ and $\varepsilon=0.9$, and (h)~$({\cal J}_{\rm NR},\rho_0)$ state diagram for $\beta_1=1.25$ and $\varepsilon=0.9$. (i--l) Hydrodynamic theory for the non-motile NRTSAIM ($\varepsilon=0$). (i)~Time evolution of $m_A$ and $m_B$ exhibiting a oscillatory state for $\beta_1=1.5$, $\rho_0=2.5$, and ${\cal J}_{\rm NR} = 0.1$. (j)~Phase portraits for $\rho_0=2.5$, ${\cal J}_{\rm NR} = 0.1$, and increasing $\beta_1$. (k)~(${\cal J}_{\rm NR}$,$\beta_1$) state diagram for $\rho_0=2.5$, and (l)~temperature-density state diagram for ${\cal J}_{\rm NR} = 0.1$.}
\label{fig9}
\end{center}
\end{figure*}

{\bf Hydrodynamic theory of the NRTSAIM.} Next, we present the results derived from the hydrodynamic theory of the NRTSAIM by solving Eqs.~\eqref{recEq5} and~\eqref{recEq6}. In this paragraph, we will keep the analytical study of these two transitions as general as possible, considering all ${\cal J}_{AB}$ and ${\cal J}_{BA}$ couplings, and the numerical solutions will be shown for ${\cal J}_{AB}=-{\cal J}_{BA}={\cal J}_{\rm NR}$.

Figs.~\ref{fig9}(a--f) show the one-dimensional steady-state density profiles of the NRTSAIM for increasing ${\cal J}_{\rm NR}$ which resemble those shown in Figs.~\ref{fig7}(g--l), obtained from simulations of the microscopic model: flocking of species A and B at small ${\cal J}_{\rm NR}$ [Figs.~\ref{fig9}(a--b)], flocking of only species A at intermediate ${\cal J}_{\rm NR}$ [Figs.~\ref{fig9}(c--d)], and a run-and-chase behavior at large ${\cal J}_{\rm NR}$ [Figs.~\ref{fig9}(e--f)]. Analogous to the microscopic simulations, the band velocity of the flocking states is larger than the self-propulsion velocity ($c/v \sim 1.17$ for ${\cal J}_{\rm NR}=0.1$), whereas the band velocity of the run-and-states is smaller ($c/v \sim 0.94$ for ${\cal J}_{\rm NR}=1$). By examining steady-state profiles for various ${\cal J}_{\rm NR}$, $\beta_1$, and $\rho_0$, we obtain qualitatively similar $({\cal J}_{\rm NR}, \beta_1)$ and $({\cal J}_{\rm NR}, \rho_0)$ state diagrams for $\varepsilon = 0.9$ in Figs.~\ref{fig9}(g--h), akin to those from simulations of the microscopic model shown in Figs.~\ref{fig8}(c--d).

The non-motile NRTSAIM, for which $\varepsilon=0$, i.e. $v=0$, exhibit only homogeneous solutions:  the Eq.~\eqref{recEq5} for densities gives $\rho_s(x,t) = \rho_0/2$ and the Eq.~\eqref{recEq6} for magnetizations becomes
\begin{align}
\frac{\dot{m_s}}{\overline{\gamma_s}} = \left( \frac{\rho_0}{2} - \frac{r_1}{2\beta_1} \right) \sinh  \frac{2\beta_1 \mu_s}{\rho_0} -m_s \cosh \frac{2\beta_1 \mu_s}{\rho_0},\label{NReq0}
\end{align}
with $m_s(x,t) = m_s(t)$, and $\mu_s =  m_s + {\cal J}_{s,-s} m_{-s}$. Stationary magnetizations are then solutions of the following coupled equations: 
\begin{equation}
m_s = \left( \frac{\rho_0}{2} - \frac{r_1}{2\beta_1} \right) \tanh \frac{2\beta_1 \mu_s}{\rho_0}. \label{eqOrder}
\end{equation}
This system of equations allows two possible solutions: a disordered homogeneous solution ($m_A=m_B=0$) and ordered homogeneous solutions ($|m_A|>0$ and $|m_B|>0$, with $m_A\ne m_B$). Due to non-reciprocal interactions, the transitions where the disordered solution becomes unstable and the homogeneous solutions are stable do not occur at the same density and temperature~\cite{vitteliNRAIM}. 

Between these two transitions, the magnetizations $m_A$ and $m_B$ exhibit an oscillatory nature [Fig.~\ref{fig9}(i)], where $m_A$ is in quadrature with $m_B$. Fig.~\ref{fig9}(j) shows the phase portraits $(m_A,m_B)$ of these stationary and non-stationary solutions as a function of $\beta_1$, for fixed $\rho_0=2.5$ and ${\cal J}_{\rm NR}=0.1$. For $\beta_1<1.4$, the disordered solution $m_A=m_B=0$ is stable. At higher $\beta_1$ ($1.4<\beta_1<1.71$), the disordered and ordered solutions are unstable, and the system exhibits a stable limit cycle. For $\beta_1>1.71$, the ordered solutions are stable, giving the emergence of four fixed points due to the symmetries of Eq.~\eqref{eqOrder}. Furthermore, the period of the oscillatory states follows the same trend as in the simulations of the microscopic model: increasing with $\beta_1$ and decreasing with ${\cal J}_{\rm NR}$ but diverging at the oscillatory/order transition.

We first derive the disorder/oscillatory transition line, corresponding to a Hopf bifurcation~\cite{vitteliNRAIM}, considering a perturbation to the disordered solution: $m_s(t) = \delta m_s(t)$. Keeping only the linear terms in $\delta m_s$ in Eq.~\eqref{NReq0}, we get
\begin{equation}
\frac{\dot{\delta m_s}}{\overline{\gamma_s}} = \left( \beta_1 - \frac{r_1}{\rho_0} \right)  \delta \mu_s  - \delta m_s,
\end{equation}
with $\delta \mu_s = \delta m_s + {\cal J}_{s,-s} \delta m_{-s}$. This equation can be rewritten in a matrix form $\dot{\delta m_s}/\overline{\gamma_s} = M_{ss'} \delta m_{s'}$, with the diagonal matrix elements $M_{ss} = \beta_1 - 1  - r_1/\rho_0$ and the off-diagonal matrix elements $M_{s,-s} = (\beta_1 - r_1/\rho_0) {\cal J}_{s,-s}$. We denote $\lambda_\pm$ the two eigenvalues of $M$, and the evolution of the perturbation is dictated by the sign of the real part of
\begin{equation}
\lambda_\pm = \beta_1 - 1  - \frac{r_1}{\rho_0} \pm \imath \left| \beta_1 - \frac{r_1}{\rho_0} \right| \sqrt{-{\cal J}_{AB} {\cal J}_{BA}}.
\end{equation}
The disordered state becomes unstable when $\beta_1 - 1 - r_1/\rho_0>0$, i.e. for a density larger than $\rho_* = r_1/(\beta_1 -1)$. This transition is then independent of ${\cal J}_{AB}$ and ${\cal J}_{BA}$ couplings, as soon as the product ${\cal J}_{AB} {\cal J}_{BA}$ is negative.

We now study the oscillatory/order transition which occurs when Eq.~\eqref{eqOrder} exhibits stable positive stationary solutions. For ${\cal J}_{\rm AB}={\cal J}_{\rm BA}=0$, i.e. with $\mu_s = m_s$, this transition occurs when $\beta_1 - r_1/\rho_0$ is larger than 1, i.e. for a density larger than $\rho_*$, telling that no oscillatory state is observed without inter-species interactions. In the presence of non-reciprocal interactions, the stability criterion for an ordered solution $(m_A,m_B)$, derived in Supplementary Note~7, reads
\begin{equation}
m_s^2 \ge m_o^2 \equiv \left( \frac{\rho_0}{2}-\frac{\rho_0+r_1}{2\beta_1} \right)\left( \frac{\rho_0}{2}-\frac{r_1}{2\beta_1} \right),
\end{equation}
giving a maximal extension for the limit circles shown in Fig~\ref{fig9}(j).

The $({\cal J}_{\rm NR},\beta_1)$ and $(T_1,\rho_0)$ state diagrams, obtained by solving numerically Eq.~\eqref{NReq0}, are shown in Figs.~\ref{fig9}(k) and~\ref{fig9}(l), respectively. The two disorder/oscillatory ($\beta_*$ or $\rho_*$) and oscillatory/order ($\beta_o$ or $\rho_o$) transition lines delimit the state diagram into three regions, revealing the presence of three distinct states: the disordered state, the oscillatory state, and the ordered state, reflecting the phase portraits shown in Fig~\ref{fig9}(j).

However, it is important to note that in our numerical simulations [Fig.~\ref{fig8}(h)], we do not observe any ordered state for any nonzero values of ${\cal J}_{\rm NR}$, even at large $\beta_1$. A similar observation is reported in the non-reciprocal Ising model~\cite{NRIM}, where the oscillatory and ordered states predicted by the mean-field description are absent, destroyed for any amount of non-reciprocity, in the 2D numerical simulations. In 3D, however, the swap state survives, likely due to reduced fluctuations, but the ordered state is eventually destroyed by non-reciprocity. In numerical simulations of our microscopic model, the oscillatory state persists [Figs.~\ref{fig8}(k--l)], likely because particles can still diffuse when $\varepsilon=0$, which stabilizes the oscillatory state.

\subsection{Metastability and motility-induced interface pinning at small diffusivity}

\begin{figure*}[t]
\begin{center}
\includegraphics[width=\textwidth]{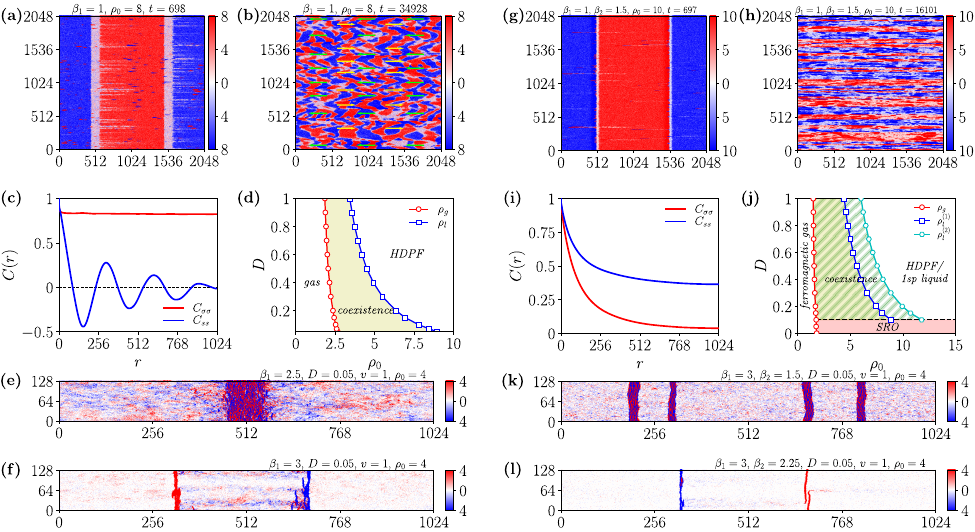}
\caption{{\it Metastability and motility-induced interface pinning for reciprocal interactions.} (a--f) Results without species flip ($\gamma_2=0$) (a)~Spontaneous nucleation of the HDPF state after 6000 MCS, for $\beta_1=1$, $D=0.05$, $v=1$ and $\rho_0=8$. (b)~Steady-state after 300000 MCS. A movie (\texttt{movie 2}) of the same can be found at Ref.~\cite{movies}. (c)~Spin-spin and species-species correlation functions along $x$: $C_{\sigma \sigma}(r) \sim \langle v_{s,i+r} v_{s,i} \rangle$ and $C_{ss}(r)\sim \langle m_i m_{i+r} \rangle$, respectively, for $\beta_1=1$, $D=0.05$, $v=1$, and $\rho_0=12$, and calculated in a $2048\times128$ domain. (d)~Diffusion-density state diagram for $\beta_1=1$ and $v=1$. (e--f) Motility-induced interface pinning for $D=0.05$, $v=1$, $\rho_0=4$, and several temperatures: (e)~$\beta_1=2.5$ and (f)~$\beta_1=3$. A movie (\texttt{movie 4}) of the same can be found at Ref.~\cite{movies}. (g--l) Results with species flip ($\gamma_2=0.5$). (g)~Spontaneous nucleation of the HDPF state after 13000 MCS, for $\beta_1=1$, $\beta_2=1.5$, $D=0.05$, $v=1$, and $\rho_0=10$. (h)~ Steady state after 300000 MCS. A movie (\texttt{movie 6}) of the same can be found at Ref.~\cite{movies}. (i)~Spin-spin and species-species correlation functions along $x$: $C_{\sigma \sigma}(r)$ and $C_{ss}(r)$, respectively, for $\beta_1=1$, $\beta_2=1.5$, $D=0.05$, $v=1$, and $\rho_0=16$, and calculated in a $2048\times128$ domain. (j)~Diffusion-density state diagram for $\beta_1=1$, $\beta_2=1.5$, and $v=1$. The SRO region exhibits steady states similar to~(h). (k--l) Motility-induced interface pinning for $\beta_1=3$, $D=0.05$, $v=1$, $\rho_0=4$, and several species coupling: (k)~$\beta_2=1.5$ and (l)~$\beta_2=2.25$. A movie (\texttt{movie 7}) of the same can be found at Ref.~\cite{movies}. For all density snapshots, the densities of A and B species are represented with red and blue colors, respectively.\label{fig10}}
\end{center}
\end{figure*}

Next, we study the fate of the liquid states of the reciprocal TSAIM without and with species flip when the diffusivity is small compared to the velocity. To do this we choose the hopping rate $W_{\rm hop}$ (defined by Eq.~\eqref{eqhop}) with $\theta=1$, i.e. $D+v$ in the favored direction $\sigma {\bf e_x}$ and $D$ in the three other directions ($-\sigma {\bf e_x}$ and $\pm {\bf e_y}$).

First, we study the fate of the liquid states of the TSAIM without species flip. Figs.~\ref{fig10}(a--b) depict the time evolution starting from the HDPF state with a diffusion constant $D = 0.05$ in a $2048 \times 2048$ domain. The bands initially move to the right ($\langle v_s \rangle >0$). After $6000$ Monte Carlo steps ($t = 698$), spontaneous nucleation of A and B droplets moving to the left appears within the high-density regions of both species [Fig.~\ref{fig10}(a)]. This spontaneous nucleation destroys the long-range order characterized by bulk HDPF {\it bands} and observed for strong diffusion ($W_{\rm hop}$ with $\theta=0$ and $D=1$), and creates a new steady state formed of alternating A and B-rich PF {\it clusters} arranged in stripes along the $y$-axis, going successively to the right and the left [Fig.~\ref{fig10}(b)]. This morphology exhibits short-range order in species but displays strong spin correlations within each stripe, unlike the one-species AIM where clusters exhibit short-range order in spin~\cite{mip}. The system is in the coexistence region, and further increasing the density does not alter the morphology but instead leads to a transition to an HDPF state. Similarly, starting from a liquid APF state, spontaneous nucleation of A and B droplets appears, and the system relaxes to the same steady state as depicted in Fig.~\ref{fig10}(b). A movie (\texttt{movie 3}) of the same can be found at Ref.~\cite{movies}.

We now analyze this new steady-state formed by the spontaneous nucleation of droplets in the HDPF state and estimate the typical sizes of the steady-state A and B clusters. Fig.~\ref{fig10}(c) shows the spin-spin and the species-species correlation functions, defined as $C_{\sigma \sigma}(r) \sim \langle v_{s,i+r} v_{s,i} \rangle$ and $C_{ss}(r) \sim \langle m_{i+r} m_i \rangle$, respectively, where $\langle \cdot \rangle$ represents spatial, temporal and ensemble averaging, and calculated in a $2048\times128$ domain. The spin-spin correlation function $C_{\sigma \sigma}(r)$ does not decay to zero along the $x$ direction, demonstrating the long-range order in the spin of the new steady state. The species-species correlation function $C_{ss}(r)$ decays to zero along the $x$ direction, and shows the structure of PF clusters in a stripe such that $C_{ss}(r) \sim \cos(\lambda_x r)\exp(-r/\xi_x)$ with a correlation length $\xi_x \simeq 128$. These pronounced oscillations of $C_{ss}(r)$, although damped, are the signature of a phase-segregated repeated pattern often observed in binary fluid~\cite{basu2016} and are also characteristic of antiferromagnetic interactions. Akin to the one-species AIM~\cite{mip}, the species-species correlation function exhibits an exponential decay along the $y$-axis with a correlation length $\xi_y \simeq 24$. These correlation functions demonstrate the short-range order in species of the new steady state and give a characteristic size to the constitutive clusters via the correlation lengths $\xi_x$ and $\xi_y$, independent of the system size.

Fig.~\ref{fig10}(d) shows the diffusion-density state diagram for $v=1$ and $\beta_1=1$, computed in the same way as the state diagrams in Figs.~\ref{fig2}(e--f) but for the hopping rate $W_{\rm hop}$ with $\theta=1$. The gas and liquid binodals $\rho_g$ and $\rho_l$ decrease with $D$, meaning one enters the {\it coexistence} region when decreasing the diffusion constant at fixed density. For a diffusion constant $D \lesssim 0.15$, the bulk HDPF bands are no longer stable and the system exhibits a steady-state morphology as depicted in Fig.~\ref{fig10}(b).

Similarly to the one-species AIM~\cite{mip}, the TSAIM exhibits motility-induced interface pinning (MIIP) at large values of $\beta_1$ (low temperatures), enhanced in the presence of small diffusion. The mechanism responsible for influencing the system's dynamics and the emergence of the MIIP transition can be attributed to the development of distinct timescales: a particle hops at a rate of $4D+v$, slow compared to the flipping, which scales as $\exp(2\beta_1)$~\cite{mip}. The MIIP state is characterized by a pinned or jammed interface, with $\sigma = +1$ particles on the left and $\sigma = -1$ particles on the right of the interface. At zero temperature, this structure is stable since a $\sigma=+1$ particle will move to the right and reach a site populated with only $\sigma=-1$ particles and then instantaneously flip to $\sigma=-1$, and return, creating a back-and-forth oscillation at the interface. Figs.~\ref{fig10}(e--f) illustrate the MIIP state for $D=0.05$ and increasing values of $\beta_1$. For $\beta_1=2.5$, the MIIP state consists of pinned interfaces formed by particles of both species, with many active particles moving outside of this structure and not contributing to the formation of the pinned interfaces [Fig.~\ref{fig10}(e)]. As we increase the inverse temperature to $\beta_1=3$, the interface pinning strengthens, resulting in the MIIP state being defined by two distinct, high-density pinned interfaces of each species, with a greater number of particles actively contributing to the formation of these interfaces [Fig.~\ref{fig10}(f)]. The MIIP states, shown in Figs.~\ref{fig10}(e--f), are independent of the initial condition (disordered, liquid APF, or HDPF state), however, with an HDPF or a liquid APF initial condition, these MIIP states can only be formed after counter-propagating droplets have spontaneously nucleated to create the pinned interface, which is enhanced when $D$ is small compared to $v$.

Similarly, we study the fate of the liquid states of the TSAIM with species flip. Figs.~\ref{fig10}(g--h) depict the time evolution starting from the HDPF state with a diffusion constant $D = 0.05$ in a $2048 \times 2048$ domain. The band movement initially goes to the right ($\langle v_s \rangle >0$). After 13000 Monte Carlo steps ($t = 697$), spontaneous nucleation of A and B droplets moving to the left appears within the high-density regions of both species [Fig.~\ref{fig10}(g)], akin to the TSAIM without species-flip.  The spin-flip mechanism governed by $W_{\rm flip}^{(1)}$ is mainly responsible for this spontaneous nucleation, which breaks the long-range order characterized by bulk HDPF {\it bands}. A new steady state is then formed by left and right flocking {\it clusters} arranged in stripes of A and B species along the $y$-axis [Fig.~\ref{fig10}(h)]. This morphology exhibits strong species correlations within each stripe but displays short-range order in spin, like the one-species AIM~\cite{mip}.

We now study this new steady-state formed by the spontaneous nucleation of droplets in the HDPF state and estimate the typical size for the steady-state left and right flocking clusters. Fig.~\ref{fig10}(i) shows the spin-spin and the species-species correlation functions, $C_{\sigma \sigma}(r)$ and $C_{ss}(r)$, and calculated in a $2048 \times 128$ domain. Akin to the one-species AIM~\cite{mip}, the spin-spin correlation function $C_{\sigma \sigma}(r)$ exhibits an exponential decay to zero with correlation lengths $\xi_x \simeq 110$ and $\xi_y \simeq 10$ along the $x$ and $y$ directions, respectively, demonstrating the short-range order in spin of the new steady state. These correlation lengths, independent of the system size, give a characteristic size to the constitutive flocking clusters. The species-species correlation function $C_{ss}(r)$ does not decay to zero along the $x$ direction and can be expressed as $C_{ss}(r) \sim C_\infty + (1-C_\infty)\exp(-r/\xi_x)$ with a correlation length $\xi_x \simeq 110$ and $C_\infty \simeq 0.38$, indicating the presence of long-range order in species within the new steady state. To emphasize, spontaneous nucleation of droplets disrupts the HDPF state in both versions of the TSAIM, with and without species flip. However, the resulting steady-state in the TSAIM without species flip exhibits long-range order in spin and short-range order in species, while the TSAIM with species flip displays the opposite behavior: long-range order in species and short-range order in spin.

Fig.~\ref{fig10}(j) shows the diffusion-density state diagram for $v=1$, $\beta_1=1$, and $\beta_2=1.5$, computed similarly to the state diagrams displayed in Figs.~\ref{fig5}(h--j) but for the hopping rate $W_{\rm hop}$ with $\theta=1$. The gas and liquid binodals $\rho_g$, $\rho_l^{(1)}$, and $\rho_l^{(2)}$ decrease with $D$. For a diffusion constant $D \lesssim 0.1$, the {\it liquid} states are no longer stable and the system exhibits a steady-state morphology as depicted in Fig.~\ref{fig10}(h), and denoted SRO (short-range ordered spin state) in this state diagram. Akin to the one-species AIM~\cite{mip}, the liquid binodals are no longer defined.

The TSAIM with species flip also exhibits MIIP at large values of $\beta_1$ (low temperatures), enhanced in the presence of small diffusion~\cite{mip}, following the same mechanism explained for the reciprocal TSAIM without species flip. Figs.~\ref{fig10}(k--l) illustrate this MIIP state for increasing values of $\beta_2$. For $\beta_2=0.75$, no MIIP state is observed and the system exhibits a ferromagnetic gas phase. For $\beta_2=1.5$, the MIIP state is formed by particles of both species, with many particles moving outside this structure and not contributing to the formation of these interfaces [Fig.~\ref{fig10}(k)]. For $\beta_2=2.25$, the MIIP state is defined by two distinct, high-density pinned interfaces of each species, due to stronger interface pinning and more particles contributing to the formation of these interfaces [Fig.~\ref{fig10}(l)]. As demonstrated, in addition to temperature $T_1$, temperature $T_2$, which governs species flip, also plays a crucial role in the emergence of the MIIP state and determining its morphology. Similarly to TSAIM without species flip, the MIIP states are independent of the initial configuration, and the formation of these interfaces is enhanced when the diffusion $D$ is small compared to the velocity~$v$.

Similarly to the reciprocal TSAIM, all the NRTSAIM steady states are metastable when the diffusion is small compared to the velocity, and transition to a disordered state. The NRTSAIM also exhibits MIIP. See Supplementary Note~8 for details.

\section{Discussion}
\label{sec7}

To summarize, we have systematically analyzed a multi-species flocking model with discrete symmetry, the two-species active Ising model (TSAIM), which serves as a discrete-symmetry counterpart of the continuous-symmetry two-species Vicsek model (TSVM)~\cite{TSVM}. Driven by recent interests~\cite{metaVM,DVM,metaTT,metaAIM,mip} in how complex and heterogeneous interactions influence active matter systems, we examine both reciprocal and non-reciprocal interactions between particles of different species, as well as the potential for a particle's species to flip from one species to another.

The flocking transition in the TSAIM with reciprocal interaction between species (without species flip) possesses many similarities with both the one-species AIM~\cite{AIMprl,AIMpre} and the TSVM~\cite{TSVM}: it exhibits a liquid-gas phase transition and displays macrophase separation in the coexistence region~\cite{AIMprl,AIMpre} where single dense liquid bands of the two species propagate on a gaseous background either in a parallel flocking (PF) state in which the two bands of the two species propagate in the same direction, or in an antiparallel flocking (APF) state in which the bands of species A and species B move in opposite directions~\cite{TSVM}. In the coexistence region, stochastic transitions between the PF and the APF states, as they can be observed in the TSVM~\cite{TSVM}, are absent in the TSAIM, and in the infinite-size limit, the system always relaxes to the PF state starting from a disordered initial condition. At large densities, the TSAIM exhibits a bistable behavior where both the HDPF and liquid APF states are stable depending on the initial condition. The liquid APF state is thermodynamically stable, corresponding to a discrete version of the TSVM liquid state, while the HDPF state emerges due to the discrete nature of the dynamics, i.e. due to normal density fluctuations, and has not been observed in previous flocking models.

The introduction of species inter-conversion in the TSAIM, which corresponds to an active extension of the equilibrium Ashkin-Teller model, further enriches the dynamical behavior of the system, leading to a wider range of steady-state phases: a spin and species-disordered paramagnetic gas phase, a spin-disordered but species-ordered ferromagnetic gas phase, a coexistence region, and a liquid state. The coexistence region can be further divided into three subregions: a ferromagnetic gas phase coexisting with a macrophase-separated liquid band, a paramagnetic gas phase coexisting with a macrophase-separated liquid band, and a paramagnetic gas phase coexisting with microphase-separated liquid bands, which have not been observed in flocking models with discrete symmetry, including the TSAIM without species flip. Moreover, the introduction of the species flip suppresses any possible APF state, which could only emerge in the scenario of a species disorder and a spin order within the liquid phase, transitioning preferentially into the microphase-separated PF state. Thus, species-flip dynamics significantly broaden the range of steady-state phases, with the interplay between initial species composition and density playing a crucial role.

Finally, our investigation of the NRTSAIM reveals the emergence of a highly efficient non-reciprocal dynamical state, termed as the {\it run-and-chase} state, when non-reciprocal frustration becomes significant. In this state, A-particles chase B-particles to align with them while B-particles avoid due to non-reciprocal interactions. This leads to a substantial accumulation of B-particles at the opposite end of the advancing A-band, allowing B-particles to maintain the maximum distance from the pursuing A-particles. This run-and-chase state can be regarded as the non-reciprocal discrete-symmetry counterpart of the chiral phase observed in the non-reciprocal Vicsek model~\cite{fruchart}. In non-reciprocal systems with discrete symmetry, the chiral phase cannot be realized due to restrictions in particle movement and reduced degrees of freedom. Our investigation further reveals that self-propulsion destroys the oscillatory state obtained for the non-motile case, as also observed in Ref.~\cite{vitteliNRAIM}.

Recent studies have argued that liquid polar flocks are metastable to the presence of small obstacles, opposite-polarity droplets, or to the spontaneous nucleation of opposite-phase droplets~\cite{metaVM,DVM,metaTT,metaAIM,mip}. In this paper, we have also confirmed that for weak diffusivity, the TSAIM long-range polar order or the NRTSAIM run-and-chase states are susceptible to spontaneous droplet nucleation, leading to a variety of steady-state system morphologies. In the TSAIM without species flip, the state is long-range ordered in spin but short-range ordered in species, whereas, in the presence of species flip, the situation is reversed. For the NRTSAIM, the resulting state is characterized by short-range ordered gas. As $\theta$ is decreased, the metastability of the HDPF state becomes less pronounced due to an increase in the nucleation time of the droplets. This is a result of the constraint $v \le 2D/(1-\theta)$, which requires large diffusion $D$ or small velocities $v$ (see  Supplementary Note~9). Moreover, at sufficiently low temperatures, all three models of TSAIM exhibit a spontaneous MIIP transition (without any impurities or disorder), as first reported in Ref.~\cite{mip}. This transition prevents the system from evolving into an HDPF state (in TSAIM) or a run-and-chase state (in NRTSAIM) when $\beta_1, \beta_2 \to \infty$.

Our study of the two-species active Ising model (TSAIM) and its non-reciprocal counterpart (NRTSAIM) provides insights into the dynamics of multi-species flocking systems. The emergence of complex steady-state phases, shaped by the interplay of diffusion, inter-species dynamics, system size, and non-reciprocal interactions, highlights their profound influence on collective behavior and emphasizes the inherent complexity of active matter systems. We hope our findings will inspire future research into the implications of these dynamics in real-world active matter systems, and further investigations into the effects of heterogeneous environments and external perturbations will provide deeper insights into the stability and adaptability of such systems.

\section{Methods}

%%%%%%%%%%%%%%%%%%%%%%%%%%%%%%%%%
%%% MODEL: SIMULATION DETAILS %%%
%%%%%%%%%%%%%%%%%%%%%%%%%%%%%%%%%

\subsection{Simulation details}

We perform Monte Carlo simulations which evolve in discrete time steps of length $\Delta t$. At each time step, $N$ randomly chosen particles are updated. A particle can either hop with a probability $p_{\rm hop} = (4D + \theta v) \Delta t$, or flip its spin orientation with a probability $p_{\rm flip}^{(1)} = W_{\rm flip}^{(1)} \Delta t \le \exp(2\beta J_1) \Delta t$, or flip its species with a probability $p_{\rm flip}^{(2)} = W_{\rm flip}^{(2)} \Delta t \le \exp(2\beta J_2) \Delta t$. The probability that nothing happens during this single particle update is $p_{\rm wait} = 1-p_{\rm hop}-p_{\rm flip}^{(1)}-p_{\rm flip}^{(2)}$. An expression for $\Delta t$ can be chosen to minimize the value of $p_{\rm wait}$:
\begin{equation}
\Delta t = \frac{1}{4D + \theta v + \gamma_1 \exp(2\beta J_1) + \gamma_2 \exp(2\beta J_2)}.
\end{equation}

This hybrid dynamics combines Monte Carlo and real-time dynamics previously used in the simulations of the one-species AIM~\cite{AIMprl,AIMpre}. We primarily consider square and rectangular domains with periodic boundary conditions in $x$ and $y$-direction to compute the steady-state density profiles and state diagrams.  We consider several initial conditions: (a) a disordered configuration with random initial positions $(x_i, y_i)$ and random spin-orientations $\sigma$; (b) a semi-ordered configuration where A and B particles are arranged in two high-density bands, either in a parallel flock (PF) state ($\sigma_A = \sigma_B$) or an anti-parallel flock (APF) state ($\sigma_A = - \sigma_B$); (c) an ordered configuration with random initial positions but with spin-orientations chosen to form a PF or APF state. Depending on the studied case, the initial population of A and B species can be either $m_0=0$ or $m_0=\rho_0$. The maximum time $t_{\rm max}$ to reach a steady state is approximately $t_{\rm max}/\Delta t \sim 10^5 - 10^7$ Monte Carlo steps (MCS). The C++ codes used for numerical simulations are available in Ref.~\cite{zenodo}.

\subsection{Numerical solutions of hydrodynamic equations}

We use the explicit Euler Forward Time Centered Space (FTCS)~\cite{ftcs} differencing scheme to numerically integrate the Eqs.~\eqref{recEq1}--\eqref{recEq6}. We solve these coupled partial differential equations on a one-dimensional ring of length $L_x$ with periodic boundary conditions, for $D=1$ and $\theta=0$, i.e. $D_{xx}=D_{yy}=1$ and $v=2\varepsilon$. In our simulation, $L_x=1024$ and the maximum simulation time is $t \simeq 10^5$. To maintain the numerical stability criteria, we set $\Delta x = 1$ and $\Delta t = 10^{-3}$ as the space and time discretizations, respectively. These discretization parameters satisfy the Courant-Friedrichs-Lewy (CFL) stability condition~\cite{cfl}. In our numerical implementation, the initial system is prepared as semi-ordered profiles with high-density regions of species A and B moving to the left or right (i.e. with positive or negative magnetization). The C++ codes used to compute the numerical solutions of Eqs.~\eqref{recEq1}--\eqref{recEq6} are available in Ref.~\cite{zenodo}.

\titleformat{\section}[block]{\centering\bfseries}{\thesection}{1pt}{}

\section*{Data Availability}
The data that support the findings of this study are available from the corresponding author upon reasonable request.

\section*{Code Availability}
The codes used in this study are available in Ref.~\cite{zenodo}.

\begin{acknowledgments}
This work was performed with financial support from the German Research Foundation (DFG) within the Collaborative Research Center SFB 1027-A3 and INST 256/539-1.
\end{acknowledgments}

\section*{Author Contributions}

MM, SC, JDN, and HR designed the research and discussed the results. MM and SC performed numerical simulations and data analysis. All authors contributed to the writing of the manuscript.

~

\section*{Competing Interests}

The authors declare no competing interests.

%\renewcommand{\thefigure}{S\arabic{figure}}
%\setcounter{figure}{0}

\begin{comment}
\appendix

\section{Nomenclature}
\label{app:nomenclature}

\begin{table}[h]
\centering
\begin{tabular}{|C{.3\linewidth}|C{.7\linewidth}|}
\hline
\textbf{Abbreviation} & \textbf{Full Form} \\ 
\hline
AIM     & Active Ising model \\
APF     & Anti-parallel flocking \\
HDPF    & High-density parallel flocking \\
MIIP    & Motility-induced interface pinning \\
NRTSAIM & Non-reciprocal two-species active Ising model \\
PF      & Parallel flocking \\
TSAIM   & Two-species active Ising model \\
TSVM    & Two-species Vicsek model \\
VM      & Vicsek model \\
\hline
\end{tabular}
\caption{List of abbreviations used in this paper and their descriptions.}
\label{tab:nomenclature}
\end{table}
\end{comment}

\clearpage

\setcounter{section}{0}
\setcounter{equation}{0}
\setcounter{figure}{0}

\renewcommand{\thesection}{Supplementary Note \arabic{section}}

\renewcommand{\thefigure}{\arabic{figure}}
\renewcommand{\figurename}{Supplementary Figure}
\renewcommand{\refname}{Supplementary References}
\renewcommand{\tablename}{Supplementary Table}
\renewcommand{\theequation}{S\arabic{equation}}

\onecolumngrid

\begin{center}
{\large \bfseries Supplemental Material for ``Emergent complex phases in a discrete flocking model with reciprocal and non-reciprocal interactions''}

\bigskip

{\normalsize Matthieu Mangeat,$^1$ Swarnajit Chatterjee,$^1$ Jae Dong Noh,$^2$ and Heiko Rieger$^1$}

\medskip

{\small \itshape $^1$Center for Biophysics \& Department for Theoretical Physics, Saarland University, 66123 Saarbr{\"u}cken, Germany.\\
$^2$Department of Physics, University of Seoul, Seoul 02504, Korea.}
\end{center}

\twocolumngrid

\titleformat{\section}[block]{\centering\bfseries}{\thesection}{1pt}{.\enspace}

\section{Derivation of hydrodynamic equations: reciprocal interactions}

The time-evolution equation for the number of particles $n_{s,i}^\sigma$ in state $\sigma$ and species $s$ on site $i$ writes
\begin{align*}
	\partial_t \langle n_{s,i}^\sigma \rangle &=   \sum_{{\bf p}}W_{\rm hop}(\sigma,{\bf p}) \left\langle n_{s,i-p}^\sigma - n_{s,i}^\sigma \right\rangle \nonumber \\
	+& \left\langle n_{s,i}^{-\sigma} W_{\rm flip}^{(1)}(-\sigma \to \sigma) - n_{s,i}^\sigma W_{\rm flip}^{(1)}(\sigma \to -\sigma) \right\rangle \nonumber \\
	+& \left\langle n_{-s,i}^{\sigma} W_{\rm flip}^{(2)}(-s \to s) - n_{s,i}^\sigma W_{\rm flip}^{(2)}(s \to -s)  \right\rangle.
\end{align*}
Noting that
\begin{equation*}
	n_{s,i-{\bf p}}^\sigma - n_{s,i}^\sigma = -{\bf p} \cdot \nabla n_{s}^\sigma + \frac{1}{2} ({\bf p} \cdot \nabla)^2 n_{s}^\sigma,
\end{equation*}
the drift term writes
\begin{align*}
	I_{\rm hop} &= \sum_{{\bf p}}W_{\rm hop}(\sigma,{\bf p}) \left[n_{s,i-{\bf p}}^\sigma - n_{s,i}^\sigma \right] \nonumber \\
	&= D(1+\theta \varepsilon) \partial_x^2 n_{s}^\sigma + D \partial_y^2 n_{s}^\sigma - 2 D \varepsilon \sigma \partial_{x} n_{s}^\sigma,
\end{align*}
using the hopping rate $W_{\rm hop}(\sigma,{\bf p})$ given by Eq.~1 of the main text. We now calculate the expression of the flipping terms:
\begin{align*}
	I_{\rm flip}^{(1)} &= n_{s}^{-\sigma} W_{\rm flip}^{(1)}(-\sigma \to \sigma) - n_{s}^{\sigma} W_{\rm flip}^{(1)}(\sigma \to -\sigma) \nonumber \\
	&= \gamma_1 \left[ n_{s}^{-\sigma} \exp\left( \frac{2 \beta_1}{\rho}  s\sigma v_a \right) - n_{s}^{\sigma}  \exp\left( -\frac{2 \beta_1}{\rho}  s\sigma v_a \right) \right],
\end{align*}
for the spin-orientation $\sigma$ (with $\beta_1 = \beta J_1$), and
\begin{align*}
	I_{\rm flip}^{(2)} &= n_{-s}^{\sigma} W_{\rm flip}^{(2)}(-s \to s) - n_{s}^{\sigma} W_{\rm flip}^{(2)}(s \to -s) \nonumber \\
	&= \gamma_2 \left[ n_{-s}^{\sigma} \exp\left( \frac{2 \beta_2}{\rho}  s m \right) - n_{s}^{\sigma}  \exp\left( -\frac{2 \beta_2}{\rho}  s m \right) \right],
\end{align*}
for the species $s$ (with $\beta_2 = \beta J_2$). These flipping terms can be rewritten as
\begin{gather*}
	I_{\rm flip}^{(1)} = - \sigma m_s \cosh \frac{2\beta_1 v_a}{\rho} + \sigma s \rho_s \sinh \frac{2\beta_1 v_a}{\rho}, \\
	I_{\rm flip}^{(2)} = - s \mu^\sigma \cosh \frac{2\beta_2 m}{\rho} + s \nu^\sigma \sinh \frac{2\beta_2 m}{\rho}, 
\end{gather*}
with $\rho_s=n_s^++n_s^-$, $m_s=n_s^+-n_s^-$, $\nu^\sigma=n_A^\sigma+n_B^\sigma$, and $\mu^\sigma=n_A^\sigma-n_B^\sigma$. The hydrodynamic equations for density $\rho = \sum_{s\sigma} \langle n_s^\sigma \rangle$, species magnetization $m = \sum_{s\sigma} s \langle n_s^\sigma \rangle$, and order parameters $v_s = \sum_{s\sigma} \sigma \langle n_s^\sigma \rangle$, and $v_a = \sum_{s\sigma} s\sigma \langle n_s^\sigma \rangle$ are:
\begin{align*}
	\partial_t \rho &= D_{xx} \partial_x^2 \rho + D_{yy} \partial_y^2 \rho - v \partial_{x} v_s, \\
	\partial_t m &= D_{xx} \partial_x^2 m + D_{yy} \partial_y^2 m - v \partial_{x} v_a \nonumber \\
	+& 2\gamma_2 \left[ \rho \sinh \frac{2\beta_2 m}{\rho} - m \cosh \frac{2\beta_2 m}{\rho} \right], \\
	\partial_t v_s &= D_{xx} \partial_x^2 v_s + D_{yy} \partial_y^2 v_s - v \partial_{x} \rho \nonumber \\ 
	+& 2 \gamma_1 \left[ m \sinh \frac{2\beta_1 v_a}{\rho} - v_s \cosh \frac{2\beta_1 v_a}{\rho} \right],\\
	\partial_t v_a &= D_{xx} \partial_x^2 v_a + D_{yy} \partial_y^2 v_a - v \partial_{x} m \nonumber \\ 
	+& 2 \gamma_1 \left[ \rho \sinh \frac{2\beta_1 v_a}{\rho} - v_a \cosh \frac{2\beta_1 v_a}{\rho} \right] \nonumber \\ 
	+& 2 \gamma_2 \left[ v_s \sinh \frac{2\beta_2 m}{\rho} - v_a \cosh \frac{2\beta_2 m}{\rho} \right],
\end{align*}
with $D_{xx}=D(1+\theta \varepsilon)$, $D_{yy}=D$, and $v=2D\varepsilon$. These equations are mean-field expressions such that $\langle f(\cdot) \rangle = f(\langle \cdot \rangle)$. As for the one-species AIM~\cite{AIMprl,AIMpre}, we must consider $m$ and $v_a$ as independent Gaussian variables with variance linear to the density $\rho$: $\sigma_m^2 = \alpha_m \rho$ and $\sigma_a^2 = \alpha_a \rho$ to observe phase-separated profiles. The hydrodynamic equations, in the refined mean-field approximation, write then:
\begin{align*}
	\partial_t \rho &= D_{xx} \partial_x^2 \rho + D_{yy} \partial_y^2 \rho - v \partial_{x} v_s, \\
	\partial_t m &= D_{xx} \partial_x^2 m + D_{yy} \partial_y^2 m - v \partial_{x} v_a \nonumber \\ 
	+& 2\overline{\gamma_2} \left[ \left(\rho- \frac{r_2}{2\beta_2} \right) \sinh \frac{2\beta_2 m}{\rho} - m \cosh\frac{2\beta_2 m}{\rho} \right],  \\
	\partial_t v_s &= D_{xx} \partial_x^2 v_s + D_{yy} \partial_y^2 v_s - v \partial_{x} \rho \nonumber \\ 
	+& 2 \overline{\gamma_1} \left[ m \sinh \frac{2\beta_1 v_a}{\rho} - v_s \cosh \frac{2\beta_1 v_a}{\rho} \right], \\
	\partial_t v_a &=  D_{xx} \partial_x^2 v_a + D_{yy} \partial_y^2 v_a - v \partial_{x} m \nonumber \\ 
	+& 2 \overline{\gamma_1} \left[ \left(\rho- \frac{r_1}{2\beta_1} \right) \sinh \frac{2\beta_1 v_a}{\rho} - v_a \cosh \frac{2\beta_1 v_a}{\rho}  \right] \nonumber \\ 
	+& 2 \overline{\gamma_2} \left[ v_s \sinh \frac{2\beta_2 m}{\rho} - v_a \cosh \frac{2\beta_2 m}{\rho} \right], 
\end{align*}
with $\overline{\gamma_i}=\gamma_i \exp(r_i/2\rho)$, $r_1=(2\beta J_1)^2 \alpha_a$, and $r_2=(2\beta J_2)^2 \alpha_m$.

\section{Derivation of hydrodynamic equations: non-reciprocal interactions}

Following the derivation made in Supplementary Note 1 for reciprocal interactions, we calculate the hydrodynamic equations for the particle density $\rho_s$, and the magnetization $m_s$ of species $s$ with non-reciprocal interactions:
\begin{align*}
	\partial_t \rho_s &= D_{xx} \partial_x^2 \rho_s + D_{yy} \partial_y^2 \rho_s - v \partial_{x} m_s, \\
	\partial_t m_s &= D_{xx} \partial_x^2 m_s + D_{yy} \partial_y^2 m_s - v \partial_{x} \rho_s \nonumber \\
	+& 2 \gamma_{1} \left[ \rho_s \sinh \frac{2\beta J_{ss'} m_{s'}}{\rho} - m_s \cosh \frac{2\beta J_{ss'} m_{s'}}{\rho}  \right],
\end{align*}
where the Einstein notation has been used: $J_{ss'} m_{s'} = J_{sA} m_A + J_{sB} m_B$. Since these equations are mean-field expressions, we must consider $m_A$ and $m_B$ as independent Gaussian variables with variance linear to the density $\rho$: $\sigma_s^2 = \alpha_s \rho$, to observe phase-separated profiles. The hydrodynamic equation for $m_s$, in the refined mean-field approximation, write then:
\begin{align*}
	\partial_t m_s &= D_{xx} \partial_x^2 m_s + D_{yy} \partial_y^2 m_s - v \partial_{x} \rho_s \nonumber \\
	+& 2 \overline{\gamma_s} \left[ \left( \rho_s - \frac{r_{ss}}{2 \beta J_{ss}} \right)\sinh \frac{2\beta_1 \mu_s}{\rho} - m_s \cosh \frac{2\beta_1 \mu_s}{\rho} \right],
\end{align*}
with $\mu_s = (J_{ss'}/J_1) m_{s'}$, $\overline{\gamma_s} = \gamma_1 \exp[(r_{ss}+r_{s,-s})/2\rho]$, and $r_{ss'} = (2\beta J_{ss'})^2 \alpha_{s'}$. We have considered that the two species have non-reciprocal interactions but are intrinsically the same, we then have $J_{AA} = J_{BB} = J_1$ and $\alpha_A = \alpha_B$. We get then $r_{AA} = r_{BB} \equiv r_1$, $r_{ss'}= (J_{ss'}/J_1)^2 r_1$, and the simplified equation:
\begin{align*}
	\partial_t m_s &= D_{xx} \partial_x^2 m_s + D_{yy} \partial_y^2 m_s - v \partial_{x} \rho_s \nonumber \\
	+& 2 \overline{\gamma_s} \left[ \left( \rho_s - \frac{r_1}{2 \beta_1} \right)\sinh \frac{2\beta_1 \mu_s}{\rho} - m_s \cosh \frac{2\beta_1 \mu_s}{\rho} \right],
\end{align*}
with $\overline{\gamma_s} = \gamma_1 \exp[(r_1+r_{s,-s})/2\rho]$.

%%%%%%%%%%%%%%%%%%%%%%%%%%%%
%%% WITHOUT SPECIES FLIP %%%
%%%%%%%%%%%%%%%%%%%%%%%%%%%%

\section{Time evolution for reciprocal interactions without species flip}

\begin{figure}[t]
\begin{center}
\includegraphics[width=\columnwidth]{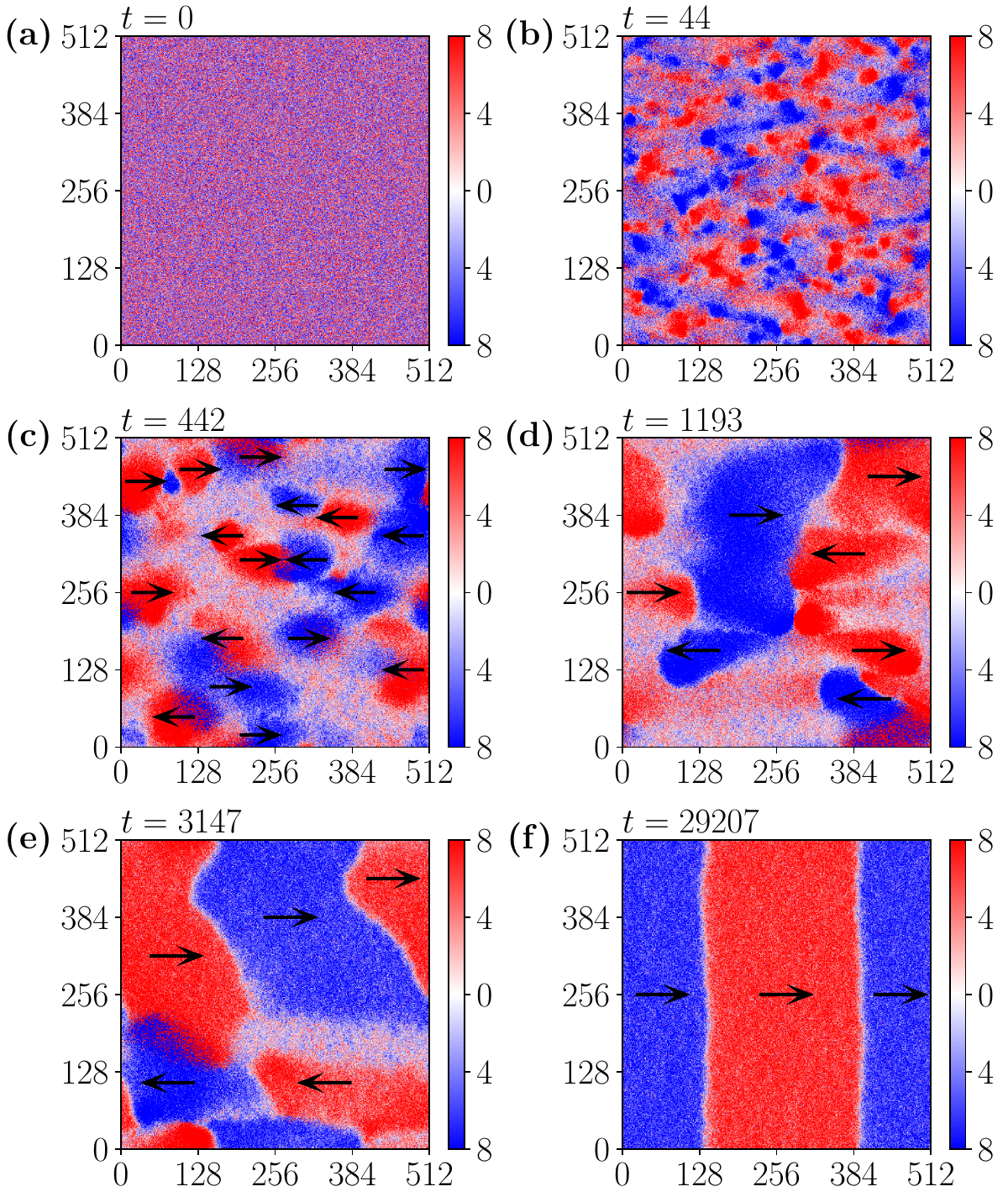}
\caption{{\it Time evolution snapshots for reciprocal interactions without species flip}. Starting from a disordered initial configuration to an HDPF state for $\beta_1=0.75$, $\varepsilon=0.9$, $\rho_0=8$ in a $512\times 512$ domain. The densities of A and B species are represented with red and blue colors, respectively. At each site, the color corresponding to the species with the higher density is displayed. \label{figS1}}
\end{center}
\end{figure}

To understand the existence of the HDPF state (observed in Fig.~1 of the main text), Supplementary Figure~\ref{figS1} here shows the time evolution of the TSAIM starting from a disordered configuration, for which the steady state is in the HDPF state. We initially observe the nucleation of several flocks of species A and B moving either to the left or right [Supplementary Figures~\ref{figS1}(a--c)]. As time progresses, these flocks grow, with pairs of parallel flocks of species A and B being favored [Supplementary Figure~\ref{figS1}(d)]. This results in the formation of multiple PF bands arranged in layers along the $y$-axis [Supplementary Figure~\ref{figS1}(e)], with alternating band motions to the left and right. At longer times, one of the layers becomes dominant, in this case moving to the right [Supplementary Figure~\ref{figS1}(f)]. Conversely, in the TSVM~\cite{TSVM}, nucleating flocks are randomly oriented, not limited to anti-parallel or parallel flocking because of the continuous symmetry of spin orientation, and the growth of these nuclei does not favor the PF state over the APF state at large times.

%%%%%%%%%%%%%%%%%%%%%%%%%
%%% WITH SPECIES FLIP %%%
%%%%%%%%%%%%%%%%%%%%%%%%%

\section{Paramagnetic and ferromagnetic gas phase for reciprocal interactions with species flip}

\begin{figure}[t]
\begin{center}
\includegraphics[width=\columnwidth]{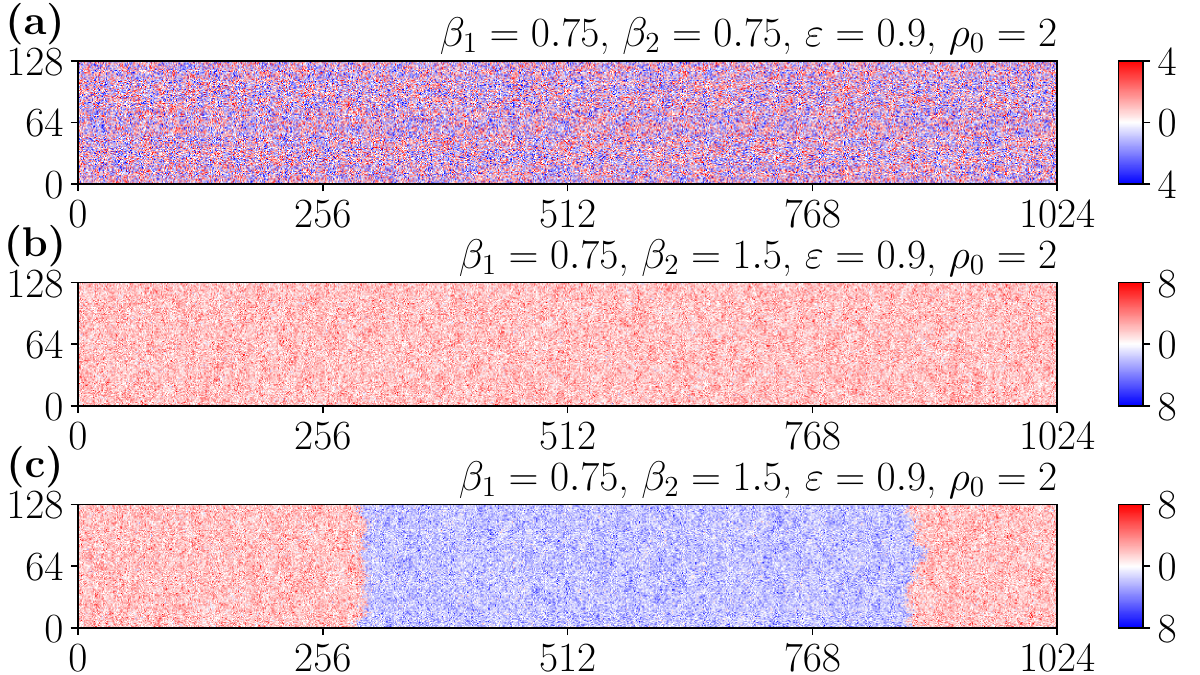}
\caption{{\it Snapshots of the different gas phases for reciprocal interactions with species flip.} (a)~Paramagnetic gas phase for $\beta_2=0.75$. (b--c)~Ferromagnetic gas phase for $\beta_2=1.5$ starting from two different initial configurations: (b)~$m_0=2$ and (c)~$m_0=0$. Red and blue colors represent the densities of A and B species, respectively. Parameters: $\beta_1=1.8$, $\varepsilon=0.9$, $\rho_0=2$. \label{figS2}}
\end{center}
\end{figure}

Supplementary Figure~\ref{figS2} shows the morphology of the different gas phases observed in the reciprocal TSAIM with species flip (see Figs.~4(a--h) in the main text) as $\beta_2$ is varied. For $\beta_2=0.75$, the particles are disordered in spin and species [Supplementary Figure~\ref{figS2}(a)], resulting in zero average local and species magnetizations, $\langle m_{s,i} \rangle = \langle m_{i} \rangle = 0$, which characterizes the {\it paramagnetic gas phase}. For $\beta_2=1.5$, the particles are disordered in spin but ordered in species [Supplementary Figures~\ref{figS2}(b--c)], resulting in zero average local magnetization, $\langle m_{s,i} \rangle = 0$, but a positive average species magnetization, $\langle m_{i} \rangle>0$, which characterizes the {\it ferromagnetic gas phase}. Starting with all particles in species A ($m_0 = \rho_0$), a ferromagnetic gas phase of species A is obtained [Supplementary Figure~\ref{figS2}(b)], whereas starting with an equal population of each species ($m_0 = 0$) distributed into two bands (semi-ordered configuration) leads to a ferromagnetic gas phase with both species occupying half of the space [Supplementary Figure~\ref{figS2}(c)].

\section{Microphase separation in the coexistence region for reciprocal interactions with species flip}

\begin{figure}[t]
\begin{center}
\includegraphics[width=\columnwidth]{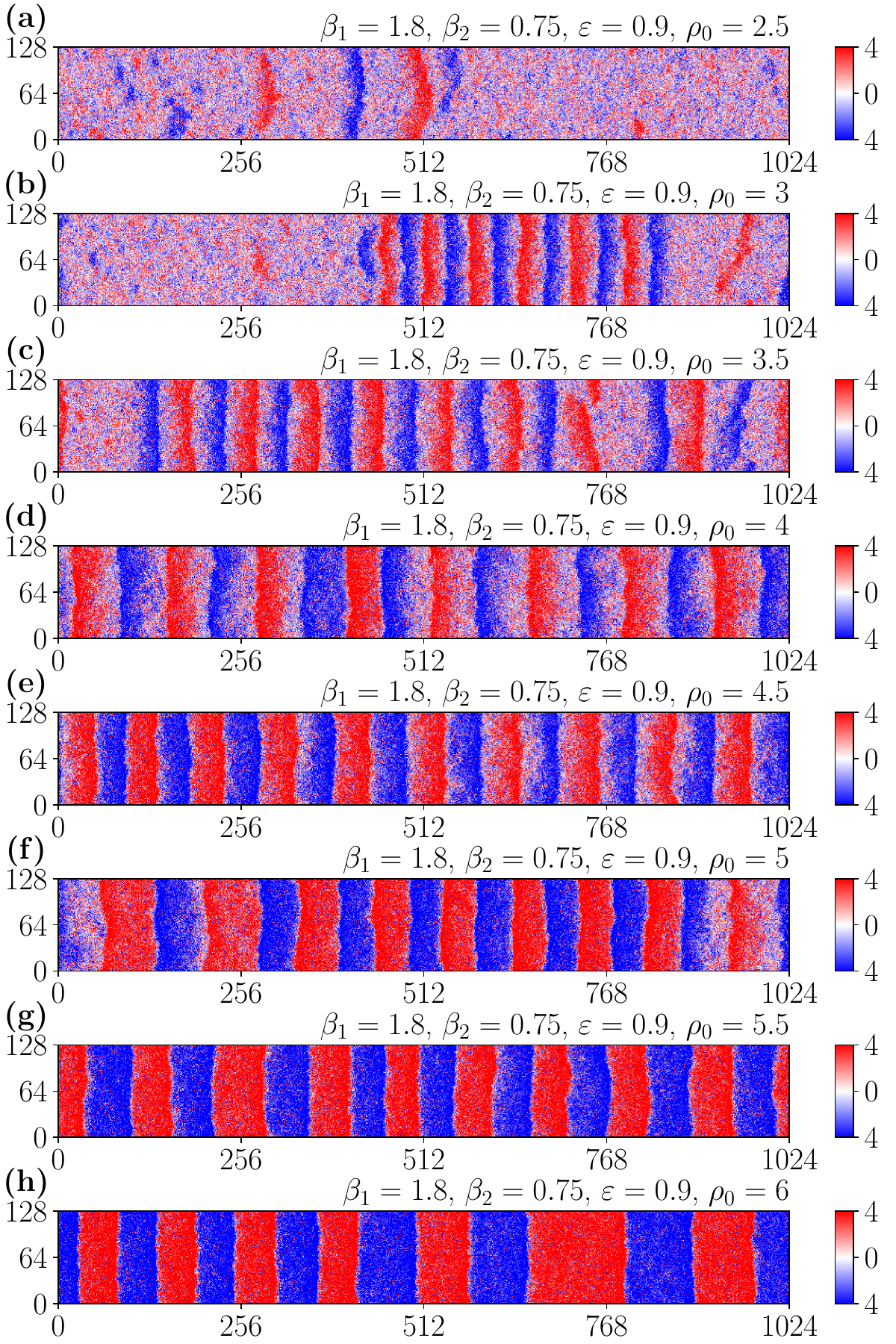}
\caption{{\it Microphase separation in the coexistence region for reciprocal interactions with species flip.} Steady-state density profiles for $\beta_1=1.8$, $\beta_2=0.75$, $\varepsilon=0.9$, and increasing density $\rho_0$, in a $1024\times128$ domain, exhibiting alternate bands of species A and B in a PF state. Red and blue colors represent the densities of A and B species, respectively. \label{figS3}}
\end{center}
\end{figure}

Supplementary Figure~\ref{figS3} shows the density profiles for species A and B as a function of increasing density for $\beta_1=1.8$ and $\beta_2=0.75$, starting from a disordered configuration. As density increases, we observe the formation of alternating dense bands of species A and B moving in a PF state, with the number and average width of these high-density bands growing as density is increased [Supplementary Figures~\ref{figS3}(a--e)]. As we further increase $\rho_0$, the number of bands initially remains constant [Supplementary Figures~\ref{figS3}(f--g)], while the average bandwidth increases before the band numbers eventually decrease as the bands become wider [Supplementary Figure~\ref{figS3}(h)]. These alternating density patterns resemble the traveling bands observed in the Vicsek model, referred to as the {\it microphase} separation of the coexistence region~\cite{solon2015a,TSVM}. However, in the Vicsek model, the microphase-separated bands exhibit a distinctive characteristic: increasing density does not alter the shape or width of these bands, instead, it only increases the number of bands. Similar patterns also emerge in the PF state of the TSVM~\cite{TSVM}, but upon increasing density, the system transitions to an APF state. In contrast, here we observe a HDPF state characterized by multiple bands. Another difference between the microphase patterns observed in Supplementary Figure~\ref{figS3} and those in the Vicsek model~\cite{solon2015a} or the TSVM~\cite{TSVM} is that, in the VM and TSVM, giant density fluctuations are responsible for the breakdown of large liquid domains. In contrast, in the TSAIM, we observe normal density fluctuations akin to the one-species AIM~\cite{AIMprl,AIMpre}, indicating that fluctuations do not play a similar role here. We should note that, for a fixed density, increasing $\beta_2$ while maintaining the condition $\beta_1 > \beta_2$ leads to the destruction of these microphase patterns. This occurs because, with a larger $\beta_2$, particles are more likely to flip their species, resulting in the formation of a single-species flock. Furthermore, the emergence of the {\it microphase} separation results from a quench starting from a disordered initial condition, and thus, for $\rho_0 > 4.5$, the formation of these multiple bands is not a steady-state property. At higher densities, if we begin with a liquid PF initial condition, the system remains in a bulk phase-separated state.

\section{Time evolution for reciprocal interactions with species flip}

\begin{figure}[t]
\begin{center}
\includegraphics[width=\columnwidth]{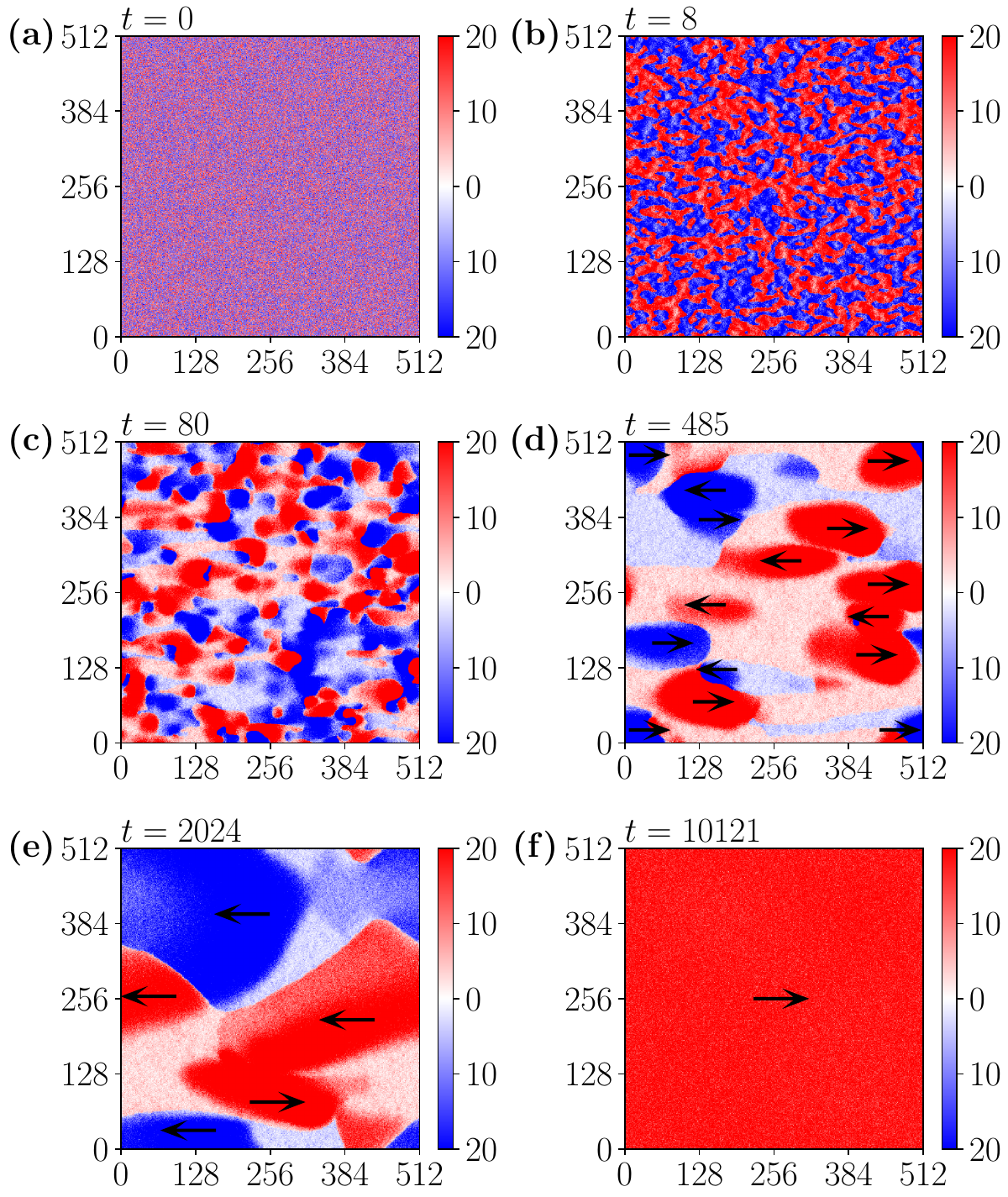}
\caption{{\it Time evolution snapshots for reciprocal interactions with species flip.} Starting from a disordered initial configuration, the system transitions to a one-species liquid phase for $\beta_1=0.75$, $\beta_2=1.5$, $\varepsilon=0.9$, $\rho_0=20$ in a $512\times 512$ domain. Red and blue colors represent the densities of A and B species, respectively. \label{figS4}}
\end{center}
\end{figure}

Supplementary Figure~\ref{figS4} shows the time evolution of the TSAIM with species flip starting from a disordered configuration (both in spin and species), for which the steady-state is in one of the liquid states (for either $m_0=0$ or $m_0=\rho_0$). Since $\beta_2$ and $\rho_0$ are large, at short times, we observe a species ordering for all the sites (decided by the sign of $m_i$), before the nucleation of several flocks of species A and B moving either to the left or right [Supplementary Figures~\ref{figS4}(a--c)]. These flocks grow with time, with a gas phase of the same species following each polar flock [Supplementary Figure~\ref{figS4}(d)]. This results in the formation of one-species flocking arranged in layers along the $y$-axis [Supplementary Figure~\ref{figS4}(e)], with alternating species (A and B) but random flocking directions. At longer times, one of the layers becomes dominant, in this case, a liquid of species A moving to the right [Supplementary Figure~\ref{figS4}(f)]. Supplementary Figure~\ref{figS4}(f), where species B is completely replaced by species A, parallels consensus formation in the binary voter model, where one opinion eventually dominates. This opinion homogenization is also known as the absorbing state (where no further changes occur) in the voter model, which typically emerges in the low-noise limit (analogous to large $\beta_2$ $(\beta_2=1.5)$ in Supplementary Figure~\ref{figS4}). The HDPF state can also be observed as a steady state, but less probable than the one-species liquid phase.

%%%%%%%%%%%%%%%%%%%%%%%%%%%
%%% WITH NR INTERACTION %%%
%%%%%%%%%%%%%%%%%%%%%%%%%%%

\section{Oscillatory/order transition line in the non-motile NRTSAIM}

\begin{figure}[t]
\begin{center}
\includegraphics[width=\columnwidth]{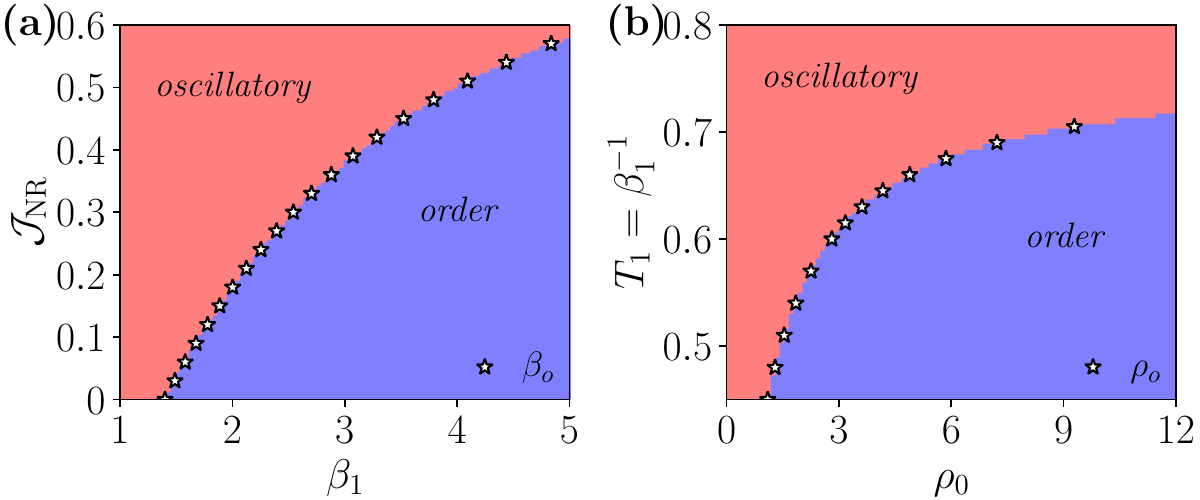}
\caption{{\it Stability diagrams of the hydrodynamics theory of the non-motile NRTSAIM.} (a)~(${\cal J}_{\rm NR}$,$\beta_1$) stability diagram for $\rho_0=2.5$, and (f)~temperature-density stability diagram for ${\cal J}_{\rm NR} = 0.1$, under the condition given by Eq.~\eqref{maxmsNR}: stable ordered solution in blue region, and unstable ordered solution in red region. The symbols report here the transition point presented in Figs.~9(k--l).\label{figS5}}
\end{center}
\end{figure}

In this section, we derive a stability criterion for an ordered solution $(m_{A,0},m_{B,0})$ following the time evolution dictated by Eq.~(21). The stationary magnetizations $m_{A,0}$ and $m_{B,0}$ are solutions of the Eq.~(22). We consider a perturbation to this ordered solution as $m_s(t)=m_{s,0} + \delta m_s(t)$. Keeping only the liner terms in $\delta m_s$ in Eq.~(21), we get
\begin{align*}
\frac{\delta \dot{m_s}}{\overline{\gamma_s}} =& \left( \beta_1 - \frac{r_1}{\rho_0} \right) \cosh (\kappa \mu_{s,0}) \delta \mu_s - \cosh (\kappa \mu_{s,0}) \delta m_s \nonumber \\
&- \kappa m_{s,0} \sinh (\kappa \mu_{s,0}) \delta \mu_s,
\end{align*}
with $\mu_{s,0} = m_{s,0} + {\cal J}_{s,-s} m_{-s,0}$, $\delta \mu_s = \delta m_s + {\cal J}_{s,-s} \delta m_{-s}$, and $\kappa=2\beta_1/\rho_0$. This equation can be rewritten in a matrix form $\dot{\delta m_s}/\overline{\gamma_s} = M_{ss'} \delta m_{s'}$ with the diagonal matrix elements:
\begin{align}
M_{ss} &= \left( \beta_1 - 1 - \frac{r_1}{\rho_0} \right) \cosh (\kappa \mu_{s,0}) - \kappa m_{s,0} \sinh (\kappa \mu_{s,0}), \nonumber \\
&=\left( \beta_1 - 1 - \frac{r_1}{\rho_0}  - \kappa^2 \frac{m_{s,0}^2}{\beta_1-r_1/\rho_0} \right) \cosh (\kappa \mu_{s,0}), \label{eqMss}
\end{align}
and the off-diagonal matrix elements ($s' \ne -s$):
\begin{align}
\frac{M_{ss'}}{{\cal J}_{ss'}} &= \left( \beta_1 - \frac{r_1}{\rho_0} \right) \cosh (\kappa \mu_{s,0}) - \kappa m_{s,0} \sinh (\kappa \mu_{s,0}), \nonumber \\
&= \left( \beta_1 - \frac{r_1}{\rho_0} \right) \left[ 1 - \left(\frac{2m_{s,0}}{\rho_0-r_1/\beta_1}\right)^2 \right] \cosh (\kappa \mu_{s,0}), \label{eqMssp}
\end{align}
where Eq.~(22) has been used to simplify the expressions. We denote $\lambda_\pm$ the two eigenvalues of $M$, and the evolution of the perturbation is dictated by the sign of the real part of
\begin{equation}
\lambda_\pm = \frac{M_{AA}+M_{BB}}{2} \pm \frac{1}{2} \sqrt{(M_{AA}-M_{BB})^2 + 4 M_{AB} M_{BA}}. \label{orderedEig}
\end{equation}
Since the oscillatory/order transition occurs at a larger density than $\rho_*$, at which the disorder/oscillation occurs, we have then $\beta_1-r_1/\rho_0 \le 1$; and Eq.~(22) tells us that $m_{s,0} < \rho_0/2-r_1/2\beta_1$. The right-hand side of Eq.~\eqref{eqMssp} is thus strictly positive, and $M_{ss'}$ has the same sign as ${\cal J}_{ss'}$, i.e. the product $M_{AB} M_{BA}$ is negative for the non-reciprocal interactions we are considering here. From the general form of the eigenvalues of $M$, given by Eq.~\eqref{orderedEig}, we get then the inequalities: $\min({M_{AA},M_{BB}}) \le \lambda_- \le \lambda_+ \le \max({M_{AA},M_{BB}})$, meaning that the ordered solution remains stable as soon as $M_{AA}$ and $M_{BB}$ are negative. From Eq.~\eqref{eqMss}, and considering $\rho_0>\rho_*$, $M_{ss}$ is negative if and only if the magnetization $m_{s,0}$ satisfies the inequality
\begin{equation}
m_{s,0}^2 \ge \left( \frac{\rho_0}{2}-\frac{\rho_0+r_1}{2\beta_1} \right)\left( \frac{\rho_0}{2}-\frac{r_1}{2\beta_1} \right).\label{maxmsNR}
\end{equation}

Supplementary Figure~\ref{figS5} shows the stability diagrams computed as follows: we consider the series of $N$ magnetization values $\{ m_{s,n}\}_{n\in\{1,N\}}$ defined by
\begin{equation*}
m_{s,n+1} = \left( \frac{\rho_0}{2} - \frac{r_1}{2\beta_1} \right) \tanh \frac{2\beta_1 \mu_{s,n}}{\rho_0},
\end{equation*}
with $\mu_{s,n} = m_{s,n} + {\cal J}_{s,-s}m_{-s,n}$ and $m_{s,1} = \rho_0/2$. If the first $N=1000$ elements satisfy the condition given in Eq.~\eqref{maxmsNR}, then the ordered solution $m_{s,0} = \lim_{n\to\infty} m_{s,n}$ is stable, otherwise no ordered solution is reached. These stability diagrams are consistent with the numerical solutions of Eq.~(21) shown in Figs.~9(k--l), and reported with symbols on Supplementary Figure~\ref{figS5}.

\section{Metastability and motility-induced interface pinning for non-reciprocal interactions}

\begin{figure*}[t]
\begin{center}
\includegraphics[width=\textwidth]{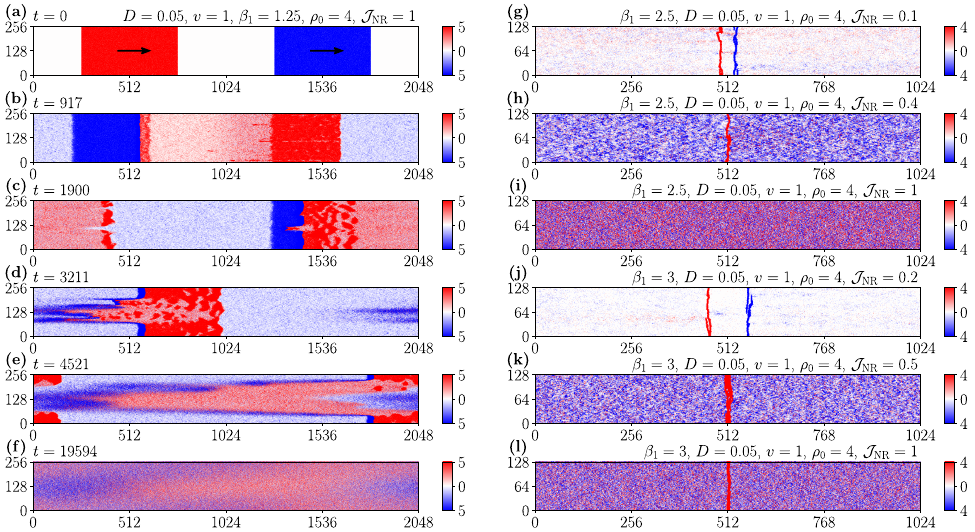}
\caption{{\it Metastability and motility-induced interface pinning for non-reciprocal interactions.} (a--f)~Time evolution from a phase-separated initial configuration, demonstrating the metastability of the run-and-chase state, for $\beta_1=1.25$, $D=0.05$, $v=1$, $\rho_0=4$, and ${\cal J}_{\rm NR}=1$, in a $2048\times256$ domain. The densities of A and B species are represented with red and blue colors, respectively. (g--l)~Motility-induced pinning in the NRTSAIM for increasing ${\cal J}_{\rm NR}$ and fixed (g--i)~$\beta_1=2.5$; and (j--l)~$\beta_1=3$. Parameters:  $D=0.05$, $v=1$, $\rho_0=4$, $L_x=1024$, and $L_y=128$. Red and
blue colors represent the densities of A and B species, respectively. A movie (\texttt{movie 10}) of the same can be found at Ref.~\cite{movies}.}\label{figS6}
\end{center}
\end{figure*}

Similarly to our analysis of reciprocal TSAIM (without and with species flip, see Fig.~10 in the main text), we now present an analysis of the stability of the NRTSAIM steady states and confirm that all these steady states are metastable due to the spontaneous nucleation of opposite spin droplets~\cite{mip}. In Supplementary Figures~\ref{figS6}(a--f), starting from an initial PF configuration where particles obey the hopping rate $W_{\rm hop}$ (defined by Eq.~1 in the main text), with $\theta=1$ and a small $D$, we demonstrate the metastability of the run-and-chase state when a droplet spontaneously nucleates in the A-band (red) [Supplementary Figures~\ref{figS6}(a--c)]. Due to the strongly non-reciprocal environment, B-particles initially accumulate away from the approaching A-particles, attempting to form a typical run-and-chase state characteristic of large ${\cal J}_{\rm NR}$. However, the spontaneous emergence of the droplet disrupts this structure [Supplementary Figures~\ref{figS6}(d--e)], causing the system to transition to a short-range gas phase [Supplementary Figure~\ref{figS6}(f)]. A movie (\texttt{movie 9}) showing the similar fate of an HDPF state in a square geometry can be found at Ref.~\cite{movies}.

Akin to one-species AIM~\cite{mip} and the reciprocal TSAIM without or with species flip (see main text), the NRTSAIM also exhibits an MIIP state at low temperature, as shown in Supplementary Figures~\ref{figS6}(g--l). At large ${\cal J}_{\rm NR}$, the MIIP clusters of species B are destroyed. Starting from an initial configuration with two high-density bands of A and B species, the system evolves under various ${\cal J}_{\rm NR}$ values. At initial times, spontaneously excited opposite polarity droplets inside the bands of each species act as the pinning sites. Consequently, the system promptly undergoes the MIIP transition for each species separately [Supplementary Figures~\ref{figS6}(g) and~\ref{figS6}(j), for ${\cal J}_{\rm NR} \le 0.3$], similar to the reciprocal TSAIM, where those small droplets gradually form multiple pinned elongated clusters that merge over time into a single, narrow, high-density elongated cluster. These clusters are jammed configurations divided by an interface: the right half consists of particles with $\sigma=-1$, and the left half consists of particles with $\sigma=+1$. When a particle crosses the interface at very low temperatures, it flips its spin state and, consequently, its self-propulsion direction, causing it to return and thereby pinning the interface. 

However, for ${\cal J}_{\rm NR} \geqslant 0.4$, the MIIP morphology observed in Supplementary Figures~\ref{figS6}(g,j) becomes unsustainable, and only the MIIP clusters of species A persist, while the MIIP clusters of species B are destroyed over time [Supplementary Figures~\ref{figS6}(h,k--l)]. As ${\cal J}_{\rm NR}$ increases, the anti-alignment strength between species B and A particles also increases. During the formation of multiple MIIP clusters for both species, a significant number of unpinned A and B particles also move in the background. Because the MIIP clusters are jammed, particle diffusion mainly occurs from the top and bottom edges of the clusters. Now, when species A particles approach a MIIP cluster of species B, a $\sigma=+1$ ($\sigma=-1$) A-particle encounters the $\sigma=+1$ ($\sigma=-1$) edges of the cluster from the left (the right). However, since species B anti-aligns with species A, the B-particle flips its spin state $\sigma$ and leaks out of the cluster (leaking from a jammed MIIP cluster occurs only when a $\sigma=+1$ particle flips to $\sigma=-1$, or vice versa). This leakage occurs at both edges, leading to the gradual reduction and eventual destruction of the B-species MIIP clusters. In contrast, when a MIIP cluster of species A is approached by species B particles, the A-particles, which prefer to align with B particles, retain their spin state and do not flip, preventing any leakage. However, when the non-reciprocal frustration is strong [compare Supplementary Figures~\ref{figS6}(i) and~\ref{figS6}(l)], the survival of the A-species MIIP cluster becomes dependent on $\beta_1$, as a large $\beta_1$ is known to facilitate pinning of the domain interface~\cite{mip}. Hence, for sufficiently large ${\cal J}_{\rm NR}$, a larger $\beta_1$ is required to observe a stable MIIP cluster.

\section{Metastability for reciprocal interactions without species flip and for $\theta<1$}

\begin{figure}[t]
\begin{center}
\includegraphics[width=\columnwidth]{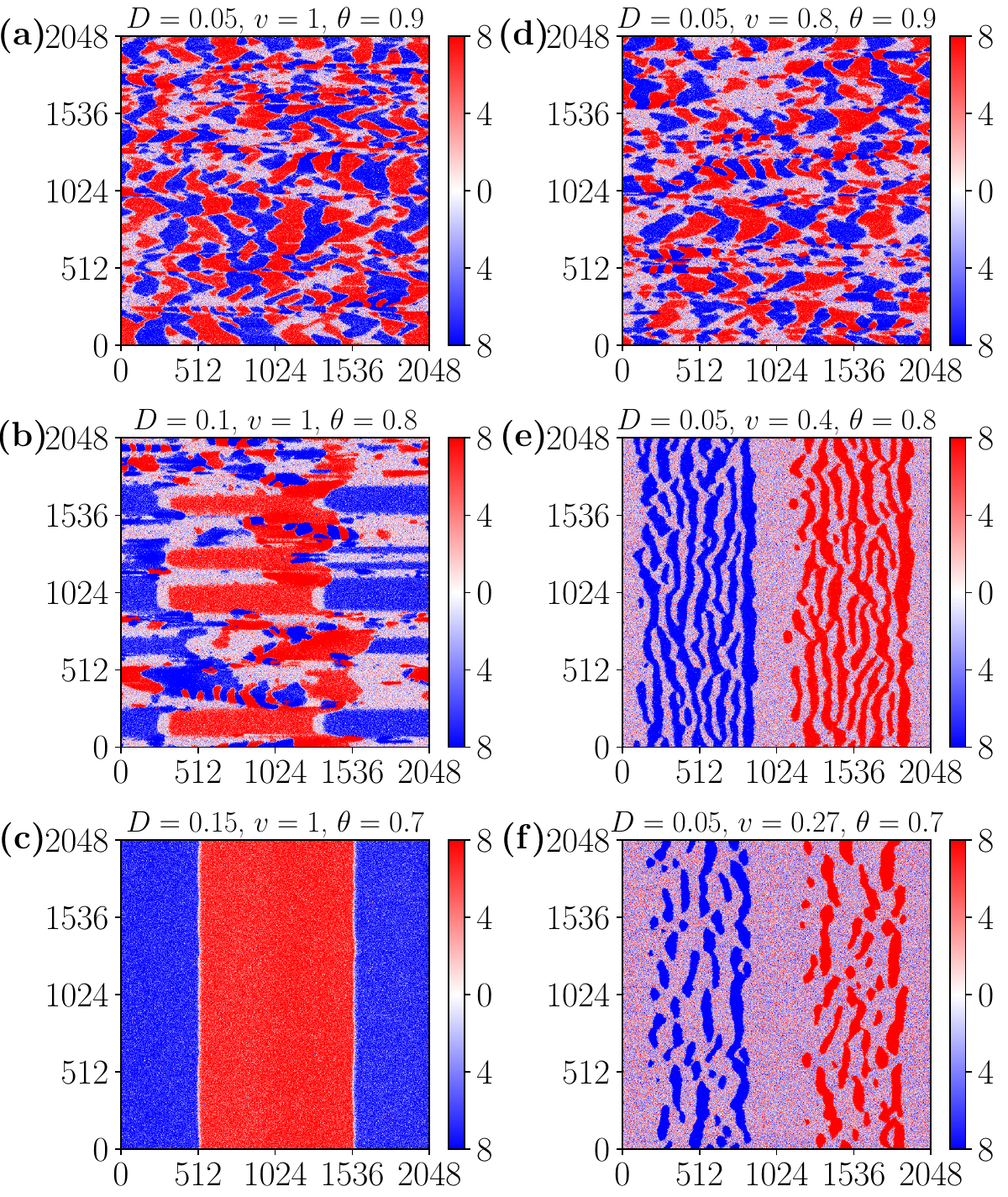}
\caption{{\it Metastability for reciprocal interactions without species flip and for $\theta<1$.} Snapshots after $1.5 \times 10^5$ MCS for $\beta_1=1$, $\rho_0=8$, and varying $\theta<1$ in a $2048\times2048$ domain, (a--c)~for $v=1$ and varying $D$, and (d--f)~for $D=0.05$ and varying $v$. The densities of A and B species are represented with red and blue colors, respectively. \label{figS7}}
\end{center}
\end{figure}

We have demonstrated that the HDPF state is vulnerable to spontaneous droplet nucleation, analogously to the one-species AIM~\cite{mip}, under the hopping rate $W_{\rm hop}$ with $\theta = 1$, resulting in a transition to a state that is long-range ordered in spin but short-range ordered in species [see Fig.~10(b)]. Supplementary Figure~\ref{figS7} shows the density profile after $t_{\rm max} = 1.5 \times 10^5$ MCS, starting from an HDPF state for varying $\theta<1$, where the inequality $v \le 2D/(1 - \theta)$ should be considered (see Sec.~II of the main text). At fixed $v=1$, the diffusion is varied as $D=(1-\theta)/2$ [Supplementary Figures~\ref{figS7}(a--c)]; while at fixed $D=0.05$, the velocity is changed as $v=0.08/(1-\theta)$ [Supplementary Figures~\ref{figS7}(d--f)]. These snapshots at fixed time $t_{\rm max}$ allow us to determine how the spontaneous droplet nucleation behaves with $\theta$. For $\theta = 0.9$, Supplementary Figures~\ref{figS7}(a,d) show a breakdown of the HDPF state leading to the formation of a new steady-state similar to $\theta=1$, shown in Supplementary Figure~10(b). For $\theta = 0.8$, the further increase of $D$ (at fixed $v=1$) makes the droplet nucleation process less frequent, and the resulting HDPF state exhibits larger correlation lengths [Supplementary Figure~\ref{figS7}(b)]. However, the decrease of $v$ (at fixed $D=0.05$) shows no spontaneous droplet excitation after a time $t_{\rm max}$, and the resulting state presents elongated liquid domains [Supplementary Figure~\ref{figS7}(e)], analogous to a quench of a liquid state into the coexistence region of the one-species AIM~\cite{AIMprl,AIMpre}. For $\theta=0.7$, the increase of $D$ significantly raises the droplet nucleation time, rendering it inaccessible after the time $t_{\rm max}$, and allowing the initial HDPF state to persist [Supplementary Figure~\ref{figS7}(c)]. Conversely, the decrease of $v$ slows down particle movement considerably, leading to fewer but much denser elongated liquid domains [Supplementary Figure~\ref{figS7}(f)]. Further decreasing of $\theta$ will require an additional increase of $D$ (for fixed $v=1$) or a decrease of $v$ (for fixed $D=0.05$), yielding an absence of spontaneous droplet nucleation after a time $t_{\rm max}$. This mainly shows that the spontaneous droplet nucleation time, increasing with $D$ and decreasing with $v$, mainly decreases with $\theta$ due to the inequality $v \le 2D/(1 - \theta)$. The influence of decreasing $\theta$ in the reciprocal TSAIM with species flip is analogous. For $\theta=0.9$, we observe a steady state characterized by short-range order in spin, similar to Fig.~10(h), while the spontaneous droplet nucleation time increases for smaller values of $\theta$.

\let\oldaddcontentsline\addcontentsline% Store \addcontentsline
\renewcommand{\addcontentsline}[3]{}% Make \addcontentsline a no-op

\let\addcontentsline\oldaddcontentsline% Restore \addcontentsline

\end{document}